\documentclass[english,11pt,aps,prd,a4paper,preprintnumbers,  floatfix,nofootinbib,showpacs,superscriptaddress,  notitlepage]{revtex4-1} 

 \pdfoutput=1
\usepackage[usenames,dvipsnames]{color}  
\usepackage{graphicx}
\usepackage{setspace}
\usepackage{upgreek}
\usepackage{braket}

\usepackage{microtype}

\usepackage{bm}
\usepackage{bbm}
\usepackage{mathrsfs}
\usepackage[bbgreekl]{mathbbol}

\usepackage{caption}
\usepackage{stackengine}
\usepackage{subcaption}
\usepackage{amsmath}
\usepackage{amssymb}
\usepackage[colorlinks=true,citecolor=darkred,urlcolor=darkred, pdfborder={0 0 0}]{hyperref}
\usepackage[normalem]{ulem}
\usepackage{xcolor}
\usepackage{graphicx}
\makeatletter
\def\p@subsection{}
\makeatother
\usepackage{xcolor}
\usepackage{float}

\definecolor{darkred}{rgb}{0.6,0,0}

\definecolor{linkcolor}{rgb}{0,0,0.5}

\def\gsim{\raise0.3ex\hbox{$\;>$\kern-0.75em\raise-1.1ex\hbox{$\sim\;$}}}
\def\lsim{\raise0.3ex\hbox{$\;<$\kern-0.75em\raise-1.1ex\hbox{$\sim\;$}}}

\def\beqn#1{\begin{equation}\label{#1}}
\def\eeqn{\end{equation}}

\def\beqa#1{\begin{eqnarray}\label{#1}}
\def\eeqa{\end{eqnarray}}

\newcommand{\mgm}{\ensuremath{\bbmu}}
\newcommand{\elm}{\ensuremath{\bbespilon}}

\usepackage{caption}
\usepackage{subcaption}
\usepackage{adjustbox}

\def\Z2{$\mathcal{Z_2}$}

\usepackage{bbold}
\usepackage{ragged2e}

\newcommand {\ignore}[1]{}

\def\cevns{CE$\nu$NS }
\def\eves{E$\nu$ES }
\usepackage{mathrsfs}

\def\321{$\mathrm{SU(3) \otimes SU(2) \otimes U(1)}$ }
\def\eves{E$\nu$ES~}

\newcommand{\AddrIISERB}{Department of Physics, Indian Institute of Science Education and Research - Bhopal, \\ 
Bhopal Bypass Road, Bhauri, Bhopal 462066, India}

\newcommand{\AddrAthens}{%
Department of Physics, National and Kapodistrian University of Athens, Zografou Campus GR-15772 Athens, Greece}

\begin{document}

\title{\textcolor{BrickRed}{Implications of first LZ and XENONnT results: A comparative study of neutrino properties and light mediators}}

\author{ShivaSankar K.A.}\email{shivasankar17@iiserb.ac.in}
\affiliation{\AddrIISERB}
\author{Anirban Majumdar}\email{anirban19@iiserb.ac.in}
\affiliation{\AddrIISERB}
\author{Dimitrios K. Papoulias}\email{dkpapoulias@phys.uoa.gr}
\affiliation{\AddrAthens}
\author{Hemant Prajapati}\email{hemant19@iiserb.ac.in}
\affiliation{\AddrIISERB}
\author{Rahul Srivastava}\email{rahul@iiserb.ac.in}
\affiliation{\AddrIISERB}

\begin{abstract}
Next generation direct dark matter detection experiments are favorable facilities to probe neutrino properties and light mediators beyond the Standard Model. We explore the implications of the recent data reported by LUX-ZEPLIN (LZ) and XENONnT collaborations on electromagnetic neutrino interactions and neutrino generalized interactions (NGIs). We show that XENONnT  places the most stringent upper limits on the effective and transition neutrino magnetic moment (of the order of few $\times 10^{-12}~\mu_B$) as well as stringent constraints to neutrino millicharge (of the order of $\sim 10^{-13}~e$)--competitive to LZ--and improved by about one order of magnitude in comparison to existing constraints coming from Borexino and TEXONO. We furthermore explore the XENONnT and LZ sensitivities to simplified models with light NGIs and find improved constraints in comparison to those extracted from Borexino-Phase II data.
\end{abstract}

\maketitle

\section{\label{sec:Intro}Introduction}
The LUX-ZEPLIN (LZ) collaboration has recently reported its first results on the search for Weakly Interacting Massive Particles (WIMPs), with a data accumulation corresponding to an exposure of $5.5$ ton for 60 live days (from 23 Dec 2021 to 11 May 2022)~\cite{LUX-ZEPLIN:2022qhg}.  At the same time, the XENONnT experiment~\citep{Aprile:2022vux}  also reported its  blind science results, collected with a total exposure of $1.16~\mathrm{ton \cdot yr}$. Interestingly, the upgraded experiment has managed to rule out the so-called XENON1T excess~\cite{XENON:2020rca} by using a new larger liquid xenon (LXe) detector with a fiducial mass of 5.9 ton and an achieved background reduction factor 5 with respect to its predecessor.   

Both experiments are based on a dual-phase LXe cylindrical time projection chamber (TPC) and have reached a very low electron recoil (ER) energy threshold of $\sim 1~\mathrm{keV_{ee}}$. This makes them ideal facilities for probing new physics phenomena involving spectral distortions at low energies. The two observables in the TPC LXe detector are the so called $S1$ and $S2$ signals, triggered by scintillation photons and subsequent ionization electrons respectively, in the aftermath of a WIMP-nucleus or background event. The first WIMP search of LZ is consistent with the null hypothesis (background-only).
The main background sources in their reconstructed ER region of interest (ROI) arise from $\beta$-decay events and elastic neutrino-electron scattering (E$\nu$ES) due to the $pp$ and $^{7}\mathrm{Be_{0.861}}$ components of the solar neutrino spectrum. Another important background source may potentially come due to $^{37}\mathrm{Ar}$ events originating from xenon exposure to cosmic rays before filling up the TPC detector and getting transferred underground. However, for the case of XENONnT this component has a negligible contribution to the background model. 

Apart from being state-of-the-art direct dark matter detection experiments--by analyzing the first LZ and XENONnT data--we show that they have also reached a better sensitivity on low-energy neutrino physics, surpassing dedicated neutrino experiments by up to an order of magnitude.  Prompted by the lack of WIMP-induced events in the ROI, in this work we are motivated to explore potential deviations from the Standard Model (SM) \eves cross section with the new data available. 
Indeed, as recently pointed out in Ref.~\cite{AtzoriCorona:2022jeb} the new LZ data can be used to set the stringent limits on effective neutrino magnetic moments,  further constraining previous limits~\cite{Coloma:2022umy} from the analysis of Borexino Phase-II data~\cite{Borexino:2017rsf} by about a factor 2.5.
These results are competitive, though slightly less stringent, to those obtained in Ref.~\cite{Khan:2022bel} using the recent XENONnT data. 
In addition to the latter works, here we also provide the corresponding constraints on the fundamental transition magnetic moments (TMMs)~\cite{Grimus:1997aa}, improving previous constraints obtained from laboratory based experiments reported in Ref.~\cite{Miranda:2019wdy}. We furthermore note that TMMs are more interesting since they have the advantage of being directly comparable with existing constraints from different laboratory experiments~\cite{AristizabalSierra:2021fuc} and astrophysics~\cite{Viaux:2013hca}.
We then demonstrate that the XENONnT and LZ data can be exploited to probe additional electromagnetic (EM) neutrino properties such as the neutrino millicharge and charge radius~\cite{Giunti:2014ixa}. Regarding LZ data, here for the first time we show that they can be used for obtaining the most severe upper limits on neutrino millicharges, which we  found to be of the order of $10^{-13}~e$.\footnote{Note, that after this study was made public, an updated version of Ref.~\cite{AtzoriCorona:2022jeb} appeared on the arXiv also exploring the sensitivity of LZ data on neutrino millicharges, where the authors found a good agreement with our present results.} We furthermore show that these sensitivities are somewhat less severe compared to those extracted from the analysis of XENONnT data, in a good agreement with Ref.~\cite{Khan:2022bel}. On the other hand, we stress that LZ and XENONnT are placing weak sensitivities on the neutrino charge radii.
We point out for the first time that the new LZ data can be used to probe neutrino generalized interactions (NGIs) due to light mediators, improving previous constraints set by Borexino. Similarly to the case of electromagnetic properties, here we confirm previous results regarding light mediator scenarios from a similar analysis performed in Ref.~\cite{Khan:2022bel} and again we find slightly improved sensitivities from the analysis of XENONnT data. Let us finally note that here we furthermore consider the case of a light tensor mediator which was neglected in Ref.~\cite{Khan:2022bel}.

The remainder of the paper is organized as follows: In Sec.~\ref{sec:theory} we discuss the \eves cross sections within and beyond the SM and we present the simulated signals expected at LZ and XENONnT.
Sec. \ref{sec:results} presents the statistical analysis we have adopted and the discussion of our results.
We finally summarize our concluding remarks in Sec. \ref{sec:conclusions}.

\section{\label{sec:theory}Theory}

\subsection{E$\nu$ES in the SM}

Within the framework of the SM the tree-level differential \eves cross section with respect to the electron recoil energy $E_{er}$, takes the form~\cite{Kayser:1979mj}
\begin{equation}
\left[\frac{d\sigma_{\nu_\alpha}}{d E_{er}}\right]_\text{SM}^{\nu e}=  \frac{G_F^2m_e}{2\pi}[(\textsl{g}_V \pm \textsl{g}_A)^2 + (\textsl{g}_V \mp \textsl{g}_A)^2\left(1-\frac{E_{er}}{E_\nu}\right)^2  -(\textsl{g}_V^2-\textsl{g}_A^2)\frac{m_e E_{er}}{E_\nu^2}] \, ,
\label{equn:EveS_SM_xsec_for_nu-alpha}
\end{equation}
where $m_e$ is the electron mass, $E_
\nu$ the incoming neutrino energy, $G_F$ the Fermi constant, while the $+ (-)$ sign accounts for neutrino (antineutrino) scattering. The vector and axial vector couplings are given by
\begin{equation}
 \label{table:EveS_SM_couplings}
 \textsl{g}_V=-\frac{1}{2}+2\sin^2\theta_W + \delta_{\alpha e}, \qquad  \textsl{g}_A=-\frac{1}{2}+ \delta_{\alpha e} \,, 
\end{equation}
with the  Kronecker delta $\delta_{\alpha e}$ term accounting for the charged-current contributions to the cross section, present only for $\nu_e$--$e^{-}$ and $\bar{\nu}_e$--$e^{-}$ scattering.

\subsection{Electromagnetic neutrino properties}
The existence of nonzero neutrino mass, established by the observation of neutrino oscillations in propagation~\cite{McDonald:2016ixn,Kajita:2016cak} stands up as the best motivation for exploring nontrivial EM neutrino properties~\cite{Schechter:1981hw, Nieves:1981zt, Kayser:1982br, Shrock:1982sc}. The most general EM neutrino vertex is expressed in terms of the EM neutrino form factors $F_q(\mathfrak{q})$, $F_\mgm(\mathfrak{q})$, $F_\elm(\mathfrak{q})$ and $F_a(\mathfrak{q})$  (for a detailed review see Ref.~\cite{Giunti:2014ixa}). The observables at a low energy scattering experiment are the charge, magnetic moment, electric moment and anapole moment, respectively, which coincide with the aforementioned EM form factors evaluated at zero momentum transfer\footnote{This is a valid approximation for solar neutrino scattering where the momentum transfer $\mathfrak{q}^2= 2 m_e E_r$ is effectively zero.} $\mathfrak{q}=0$.

The helicity-flipping neutrino magnetic moment contribution to the \eves cross section adds incoherently to the SM and reads~\cite{Vogel:1989iv}
\begin{equation}
\label{equn:EveS_mag_xsec_for_nu-alpha}
\begin{aligned}
\left[\frac{d\sigma_{\nu_\alpha}}{d E_{er}}\right]_\text{mag}^{\nu e}=&\frac{\pi a^2_\text{EM}}{m_e^2}\left[\frac{1}{E_r} - \frac{1}{E_\nu}\right] \left( \frac{\mu^\text{eff}_{\nu_\alpha}}{\mu_B} \right)^2\, ,
\end{aligned}
\end{equation}
with $a_\text{EM}$ denoting the fine structure constant. Note that the so-called effective magnetic moment is expressed in terms of the fundamental neutrino magnetic ($\mgm$) and electric ($\elm$) dipole moments, which for solar neutrinos takes the form $\mu^\text{eff}_{\nu_\alpha} =\sum_k \vert \sum_j U^*_{\alpha k} \lambda_{jk} \vert^2$, where $\lambda_{jk} =  \mgm_{jk} - i \elm_{jk}$ represent the TMMs~\cite{Grimus:1997aa, AristizabalSierra:2021fuc}. 
For Majorana neutrinos, the latter is an antisymmetric matrix which in the mass basis takes the form~\cite{Grimus:2002vb}
 \begin{equation}
 \lambda = \left( \begin{array}{ccc}
 0 & \Lambda_3 & - \Lambda_2 \\
 - \Lambda_3 &  0 & \Lambda_1 \\
 \Lambda_2 & - \Lambda_1 & 0
 \end{array} \right) \, , 
 \label{NMM:matrix}
 \end{equation}
 where for simplicity the definition $\Lambda_{i} = \epsilon_{ijk} \lambda_{jk}$ has been introduced. Then, the most general effective neutrino magnetic moment taking into account also the effect of neutrino oscillations in propagation reads~\cite{Beacom:1999wx}
 \begin{equation}
 \mu_{\nu,\text{eff}}^2 (L, E_\nu) = \sum_j \Big \vert \sum_i U^\ast_{\alpha i} e^{-i\, \Delta m^2_{ij} L /2 E_\nu} \lambda_{ij} \Big \vert^2 \, ,
 \label{NMM-observable}
 \end{equation}
 with $U_{\alpha i}$ being the entries of the lepton mixing matrix, $\Delta m^2_{ij}$ the neutrino mass splittings and $L$ the distance between the neutrino source and detection points. For the case of solar neutrinos we are interested in this work, the exponential in Eq.(\ref{NMM-observable}) can be safely neglected and the effective neutrino magnetic moment takes the form~\cite{Miranda:2020kwy}
 \begin{equation}\label{eq:nmm_sun}
  (\mu^\text{eff}_{\nu,\,\text{sol}})^{2} = |\mathbf{\Lambda}|^{2} - 
   c^{2}_{13}|\Lambda_{2}|^{2} + (c^{2}_{13}-1)|\Lambda_{3}|^{2} +
   c^{2}_{13}P^{2\nu}_{e1}(|\Lambda_{2}|^{2}-|\Lambda_{1}|^{2})\, ,
 \end{equation}
 with $|\mathbf{\Lambda}|^{2}=|\Lambda_{1}|^{2}+|\Lambda_{2}|^{2}+|\Lambda_{3}|^{2}$, $c_{13} = \cos\theta_{13}$ and $P_{e1}^{2\nu} = 0.667 \pm 0.017$ which corresponds to the average probability value for $pp$ neutrinos~\cite{deSalas:2020pgw}. The latter expression is used to map between the effective neutrino magnetic moments and the fundamental TMMs.

On the other hand the  helicity-preserving EM contributions for millicharge $(q_{\nu_\alpha})$, anapole moment $(a_{\nu_\alpha})$ and neutrino charge radius\footnote{Even for vanishing neutrino charge, neutrino charge radii can be generated from radiative corrections through the term $\langle r^2_\nu \rangle = 6 \, d F_q(\mathfrak{q^2})/\mathfrak{q^2}\vert_{\mathfrak{q =0}}$.} $(\langle r^2_{\nu_\alpha} \rangle)$ are taken via the substitution~\cite{Giunti:2014ixa}:
\begin{equation}
\label{Equn:CR_Anapole_FC_xSec}
g_V\rightarrow g_V+\frac{\sqrt{2}\pi a_\text{EM}}{G_F}\left[\frac{\left<r_{\nu_\alpha}^{2}\right>}{3}-\frac{a_{\nu_\alpha}}{18}-\frac{1}{m_eE_{er}}\left(\frac{q_{\nu_\alpha}}{e}\right)\right]\, ,
\end{equation}
where $e$ is the electric charge of electron.

\subsection{Light mediators}
Sensitive experiments with extremely low-energy threshold capabilities such as LZ and XENONnT constitute excellent probes of new physics interactions that involve spectral features induced in the presence of novel mediators~\cite{Ge:2020jfn, Ge:2021snv}. Many such beyond the SM physics scenarios can be accommodated in the context of model independent NGIs~\cite{Lindner:2016wff, Rodejohann:2017vup}. Let us note that in this framework all Lorentz invariant forms $X=\{S,P,V,A,T\}$ employing Wilson coefficients of dimension-six effective operators can be incorporated~\cite{AristizabalSierra:2018eqm}. Here, we consider the \eves contributions of light scalar ($S$), pseudoscalar ($P$), vector ($V$), axial vector ($A$) and tensor ($T$) bosons with mass $m_X$ and coupling $\textsl{g}_X= \sqrt{\textsl{g}_{\nu X} \textsl{g}_{eX}}$, and explore how well they can be constrained in the light of the recent data. For $X = \{V,A\}$ interactions, the corresponding differential cross sections can be obtained from Eq.(\ref{equn:EveS_SM_xsec_for_nu-alpha}) and the replacements ~\cite{Lindner:2018kjo} 
\begin{equation}
\label{equn:EveS_BSM_xsec_V,A}
\textsl{g}'_{V/A}=\textsl{g}_{V/A}+\frac{\textsl{g}_{\nu V/A}\cdot \textsl{g}_{e V/A}}{2\sqrt{2}G_F(2m_e E_{er} + m_{V/A}^2)} \, .
\end{equation}
For the case of $X = \{S, P, T\}$ interaction there is no interference with the SM cross section and the relevant contributions read~\cite{Link:2019pbm}
\begin{equation}
\left[\frac{d\sigma_{\nu_\alpha}}{d E_{er}}\right]_{S}^{\nu e}=\left[\frac{\textsl{g}^2_{\nu S}\cdot \textsl{g}^2_{eS}}{4\pi(2m_e E_{er} + m_{S}^2)^2}\right]\frac{m_e^2 E_{er}}{E_\nu^2}\, ,
\end{equation}
\begin{equation}
\left[\frac{d\sigma_{\nu_\alpha}}{d E_{er}}\right]_{P}^{\nu e}= \left[\frac{\textsl{g}^2_{\nu P}\cdot \textsl{g}^2_{eP}}{8\pi(2m_e E_{er} + m_{P}^2)^2} \right]\frac{m_e E_{er}^2}{E_\nu^2}\, ,
\end{equation}
\begin{equation}
\label{equn:EvES_Tensor}
\left[\frac{d\sigma_{\nu_\alpha}}{d E_{er}}\right]_{T}^{\nu e}= \frac{m_e\cdot \textsl{g}^2_{\nu T}\cdot \textsl{g}^2_{eT}}{\pi(2m_e E_{er} + m_{T}^2)^2}\cdot \Bigl[1+2\left(1-\frac{E_{er}}{E_\nu}\right) + \left(1-\frac{E_{er}}{E_\nu}\right)^2-\frac{m_e E_{er}}{E_\nu^2}\Bigr]\, .
\end{equation}

\subsection{Simulated event rates at LZ and XENONnT}
At LZ and XENONnT the differential \eves event rate for the different interactions $\xi=\{\mathrm{SM,~EM,~NGI}\}$ is calculated, as~\cite{AristizabalSierra:2017joc}
\begin{equation}
\left[\frac{dR}{dE_{er}}\right]_{\xi}^{\nu e} = \mathcal{E} N_\text{T}\sum_{i = \mathrm{solar}}\int_{E_\nu^\text{min}}^{E_\nu^\text{max}}dE_{\nu}\frac{d\Phi^{\nu}_i(E_\nu)}{dE_\nu} \left[\frac{d\sigma}{d E_{er}}\right]_{\xi}^{\nu e} 
\label{equn:EvES_Differential_Event_Rate}
\end{equation}
where $\mathcal{E}$ and $N_T=Z_\text{eff}m_\text{det}N_A/m_\text{Xe}$ denote the exposure and number of electron targets respectively, with $m_\text{det}$ being the fiducial mass of the detector, $N_A$  the Avogadro number and $m_\text{Xe}$ the molar mass of $^{131}\text{Xe}$. Due to atomic binding, $Z_\text{eff}(E_{er})$ accounts for the number of electrons that can be ionized for an energy deposition $E_{er}$. The latter is approximated through a series of step functions that depend on the single particle  binding energy of the $i$th electron, evaluated from Hartree-Fock calculations~\cite{Chen:2016eab}. In the ROI of LZ and XENONnT experiments, \eves populations are mainly due to $pp$ neutrinos with a subdominant contribution coming from $^{7}\text{Be}_{0.861}$ neutrinos, while the rest fluxes of the solar neutrino spectrum~\cite{Baxter:2021pqo}, $(d\Phi^{\nu}_i/dE_\nu)$, contribute negligibly. Since solar neutrinos undergo oscillations in propagation before reaching the Earth, the cross section in Eq.(\ref{equn:EvES_Differential_Event_Rate}) is weighted with the relevant oscillation probability and reads
\begin{equation}
\left[\frac{d\sigma}{d E_{er}}\right]_{\xi}^{\nu e} = P_{ee} \left[\frac{d\sigma_{\nu_e}}{d E_{er}}\right]_{\xi}^{\nu e} + \sum_{f=\mu, \tau} P_{ef} \left[\frac{d\sigma_{\nu_f}}{d E_{er}}\right]_{\xi}^{\nu e} \, ,
\label{eq:xsec_flavored}
\end{equation}
where $P_{ee} = \cos^4 \theta_{13} \mathcal{P}_\text{eff}+ \sin^4 \theta_{13} $ is the  solar neutrino survival probability calculated in the two-flavor approximation following the standard procedure detailed in Ref.~\cite{AristizabalSierra:2017joc}, while $P_{e \mu}=(1-P_{ee}) \cos^2{\theta_{23}} $ and $P_{e \tau}=(1-P_{ee}) \sin^2{\theta_{23}} $ are the corresponding transition probabilities with the atmospheric mixing angle $\theta_{23}$ taken from~\cite{deSalas:2020pgw}. Here, $\mathcal{P}_\text{eff}= \left(1 + \cos 2 \theta_M \cos 2 \theta_{12} \right)/ 2 $ which depends also on the neutrino propagation path and accounts for matter effects~\cite{Parke:1986jy} with neutrino production distribution functions and neutrino fluxes (pp and CNO). For evaluating $\mathcal{P}_\text{eff}$, the required oscillation parameters $\Delta m^2_{12}, \theta_{12}, \theta_{13}$ are all taken from Ref.~\cite{deSalas:2020pgw} assuming their central values for normal ordering. The minimum neutrino energy required to induce an electronic recoil $E_{er}$ is trivially obtained from the kinematics of the process, as $
 E_\nu^\text{min}=\left(E_{er}+\sqrt{E_{er}^2+2m_e E_{er}}\right)/2$.

\begin{figure}[t]
 \includegraphics[width=0.49 \textwidth]{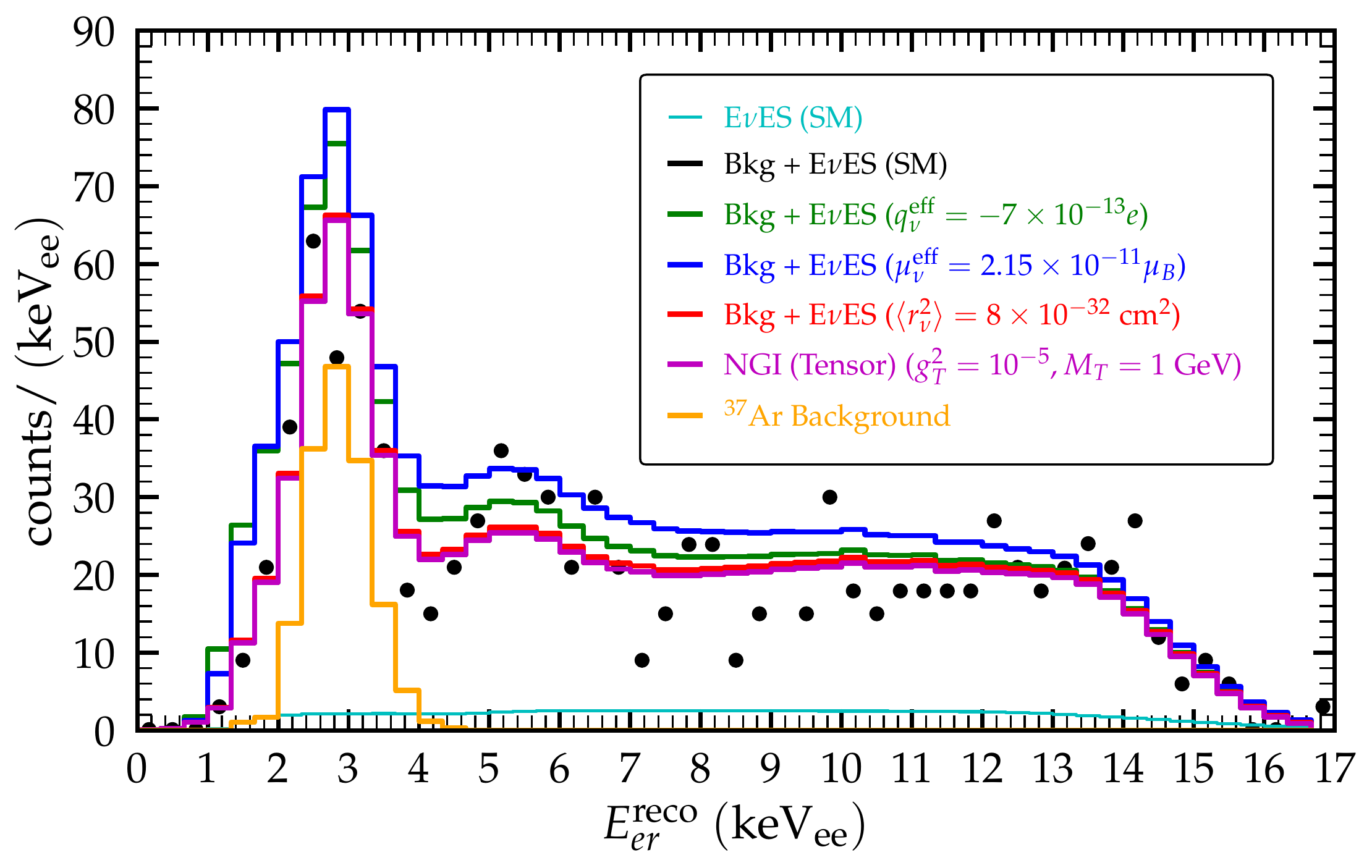}
 \includegraphics[width=0.49 \textwidth]{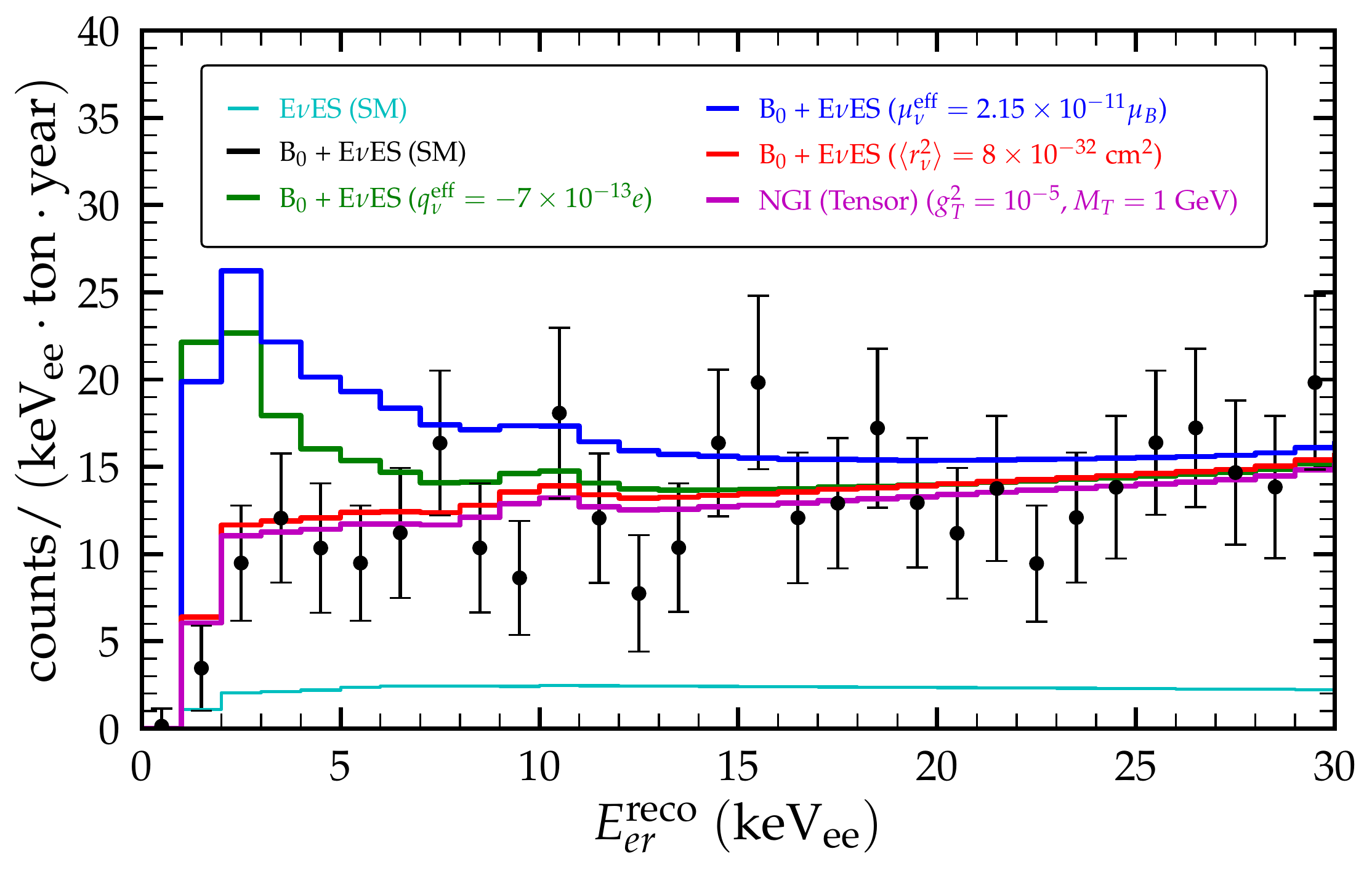}
 \caption{Expected signal and comparison with the experimental data at LZ (left) and XENONnT (right). Various examples of possible new physics contributions to \eves are shown.}
 \label{fig:events}
 \end{figure}

In order to accurately simulate the LZ signal, the true differential event rate of Eq.(\ref{equn:EvES_Differential_Event_Rate}) is then smeared with a normalized Gaussian resolution function with width $\sigma(E_{er}^\text{reco})=K\cdot\sqrt{E_{er}^\text{reco}}$~\cite{Pereira} ($E_{er}^\text{reco}$  is  the reconstructed recoil energy and $K=0.323\pm 0.001~\mathrm{(keV_{ee})}^{1/2}$) and converted to a reconstructed spectrum.
Finally, the reconstructed spectrum is weighted by the efficiency function, $\mathcal{A}(E_{er}^\text{reco})$ taken from the  Ref.~\cite{AtzoriCorona:2022jeb}\footnote{The ER efficiency is taken from arXiv V3 of Ref. \cite{AtzoriCorona:2022jeb}.}, where the authors used the NEST 2.3.7~\cite{szydagis_m_2022_6534007} software and the LZ data release~\cite{LUX-ZEPLIN:2022qhg} for accurately extracting the efficiency in terms of ER, since in the data release that latter was provided  in units of nuclear recoil energy. For the case of XENONnT, a similar procedure is followed, while the efficiency is taken from Ref.~\cite{Aprile:2022vux} and the resolution function from Ref.~\cite{XENON:2020rca}. In Fig.~\ref{fig:events} we present a comparison of the experimental data with the expected signal at LZ (left panel) and XENONnT (right panel) for various new physics scenarios involving EM neutrino properties and NGIs. We have furthermore checked that our integrated SM \eves spectra in the ROI of LZ and XENONnT agree well with those reported by the two collaborations.
%

\section{\label{sec:results}Results and Discussion}

In order to explore the new physics parameter(s) of interest $\mathcal{S}$ with the LZ data, our statistical analysis is based on the Poissonian $\chi^2$ function~\cite{Feldman:1997qc}
\begin{equation}
\chi^2(\mathcal{S})=2\sum_{i=1}^{51} \Biggl[R_\mathrm{pred}^i(\mathcal{S})- R_\mathrm{exp}^i+  R_\mathrm{exp}^i\ln \left(\frac{R_\mathrm{exp}^i}{R_\mathrm{pred}^i(\mathcal{S})}\right) \Biggr]+\left(\frac{\alpha}{\sigma_\alpha}\right)^2+\left(\frac{\beta}{\sigma_\beta}\right)^2+\left(\frac{\delta}{\sigma_\delta}\right)^2 \, ,
\label{equn:chi_square}
\end{equation}
where $R_\mathrm{exp}^i$ stands for the experimental differential events in $i$th bin reported in~\cite{LUX-ZEPLIN:2022qhg}, while the predicted differential spectrum--which contains \eves and background components-- is taken as $R_\mathrm{pred}^i(\mathcal{S}; \alpha,\beta,\delta)=(1+\alpha)R_\mathrm{bkg}^i+(1+\beta)R_\text{\eves}^i(\mathcal{S})+(1+\delta)R_{^{37}\text{\text{Ar}}}^i$. Here, the nuisance parameters  $\alpha$, $\beta$ and $\delta$ are introduced to incorporate the uncertainty on background, flux normalization and $^{37}\text{Ar}$ components with $\sigma_\alpha=13\%$, $\sigma_\beta=7\%$ and $\sigma_\delta=100\%$ (see Refs.~\cite{LUX-ZEPLIN:2022qhg,AtzoriCorona:2022jeb}). Let us note that while we do not vary the oscillation parameters in our fitting procedure, their uncertainty is effectively accounted for in the large flux normalization uncertainty.  Following Ref.~\cite{AtzoriCorona:2022jeb}, the $R_\mathrm{bkg}^i$ spectrum is taken by subtracting the SM and $^{37}\text{\text{Ar}}$ contributions from the total background reported in~\cite{LUX-ZEPLIN:2022qhg} by normalizing the  integrated spectrum of $^{37}$Ar to its nominal value given by the LZ collaboration, i.e. 97 events.
For the case of XENONnT we employ a Gaussian $\chi^2$ function
\begin{equation}
\chi^2(\mathcal{S})= \sum_{i=1}^{30} \left( \frac{R_\text{pred}^i(\mathcal{S},\beta)-R_\text{exp}^i}{\sigma^i} \right)^2 + \left( \frac{\beta}{\sigma_\beta}\right)^2 \, ,
\end{equation}
where $R_\text{pred}^i(\mathcal{S},\beta)=\left(1+\beta \right) R^{i}_\text{\eves}(\mathcal{S}) + B_0^i$. Here, $B_0$ represents the modeled background reported in~\cite{Aprile:2022vux} from which we have subtracted the SM \eves contribution.

\begin{figure}[t]
\begin{center}
\includegraphics[width=0.49\textwidth]{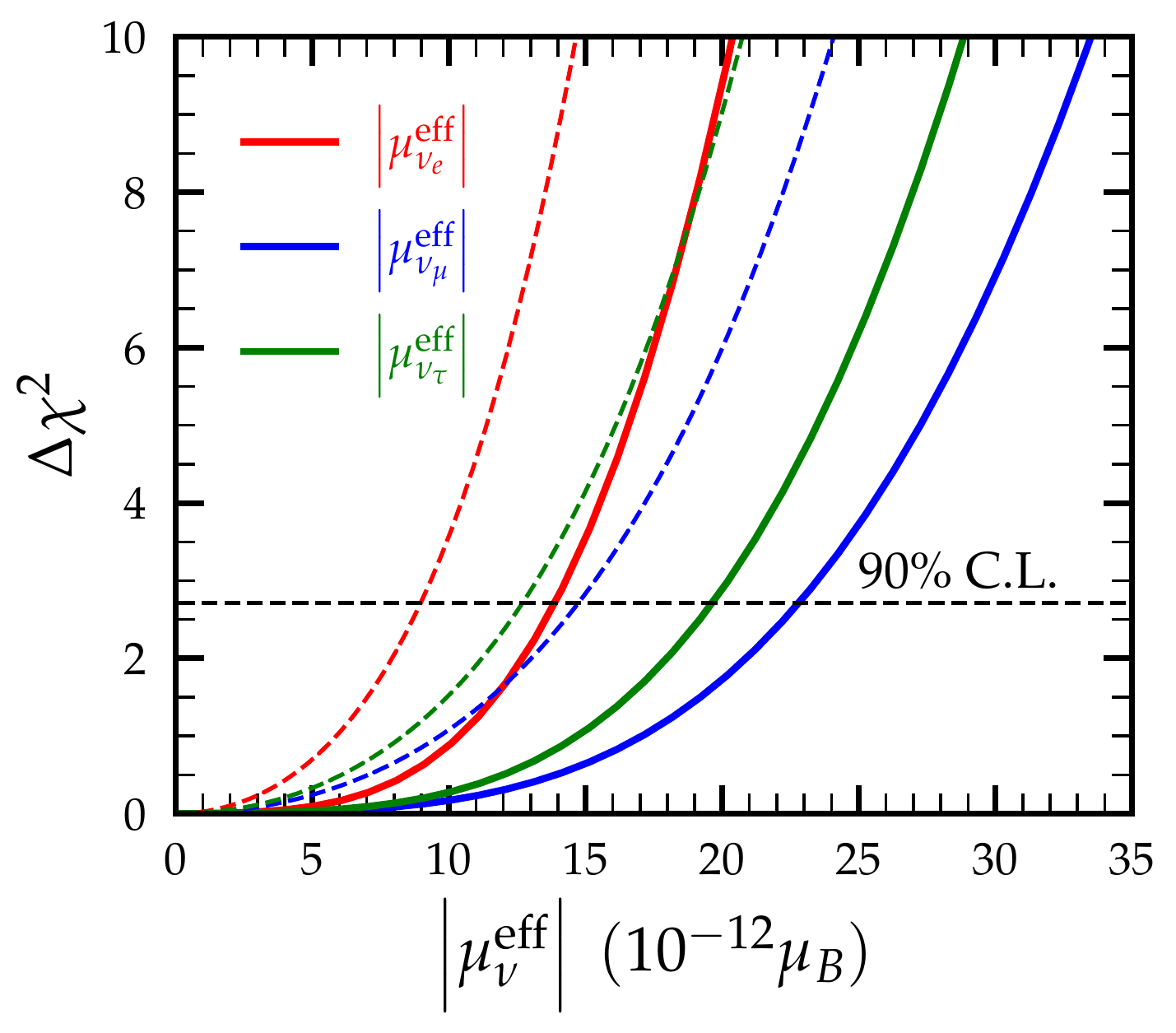}
\end{center}
\caption{$\Delta\chi^2$ profiles of the flavor dependent effective neutrino magnetic moment. The results correspond to the analysis of LZ data (solid lines) and XENONnT data (dashed lines).}
\label{fig:v_Electro_Magnetic_Properties}
\end{figure}

\begin{figure*}
 \includegraphics[width=0.32 \textwidth]{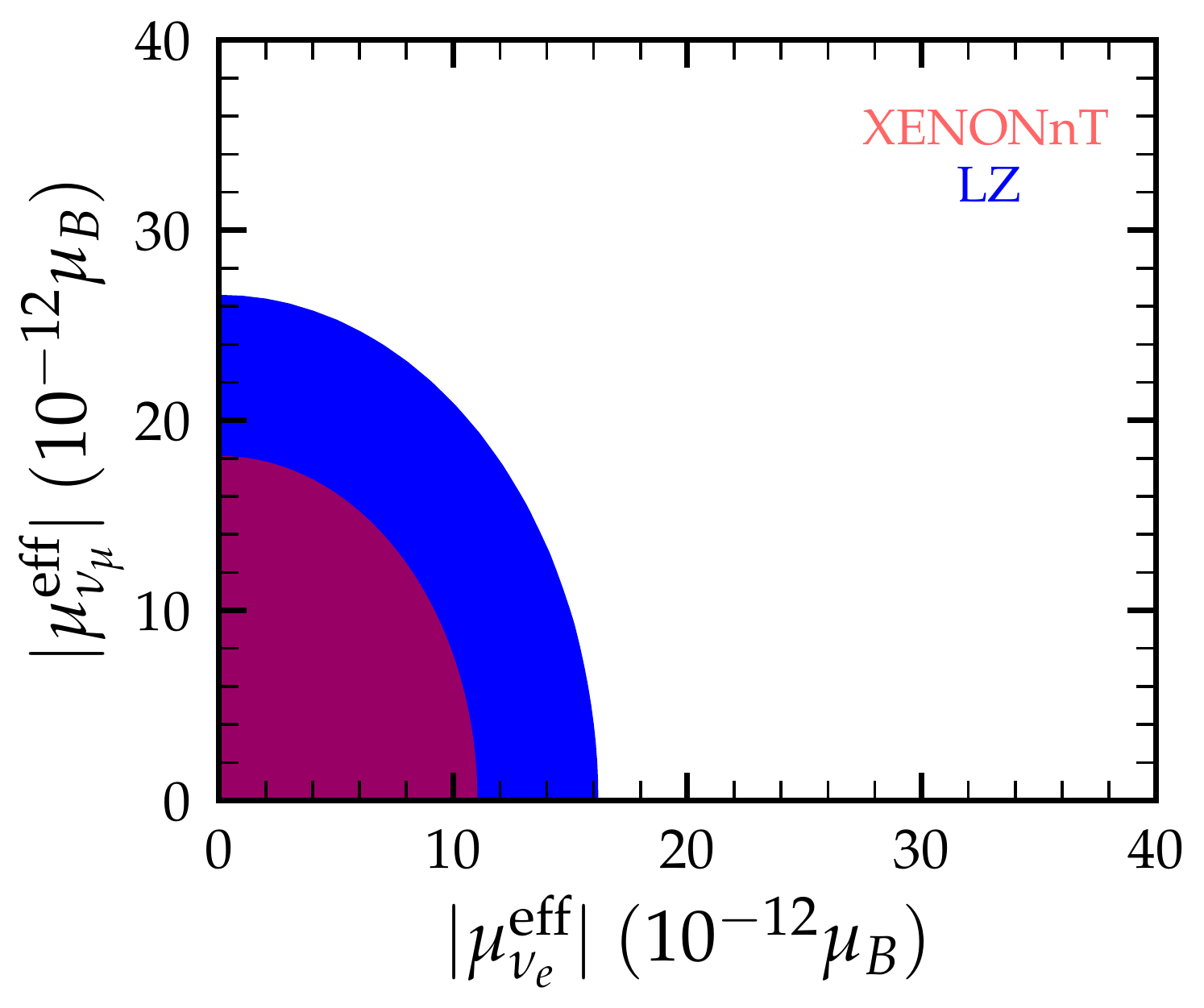}
 \includegraphics[width=0.32 \textwidth]{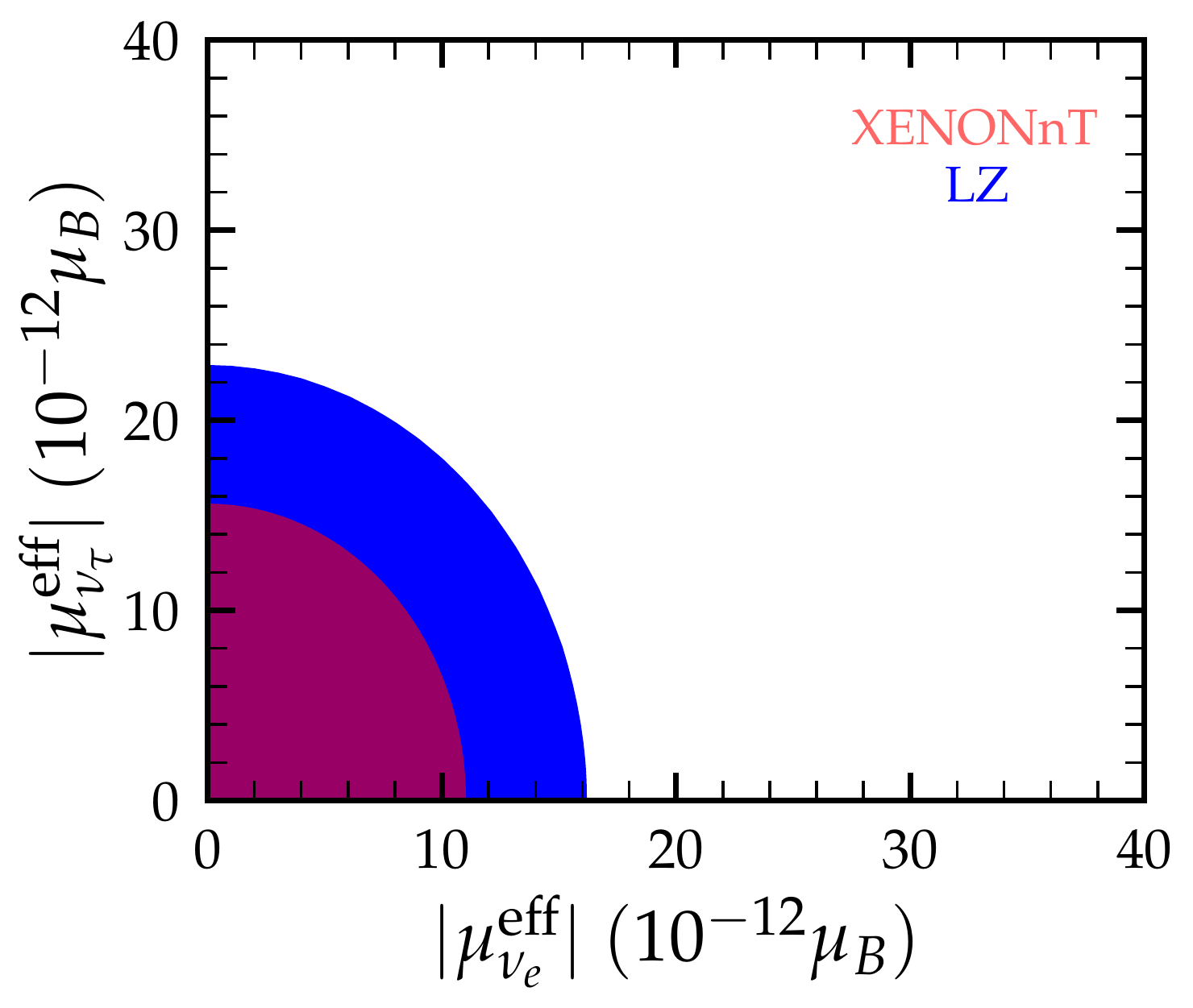}
 \includegraphics[width=0.32 \textwidth]{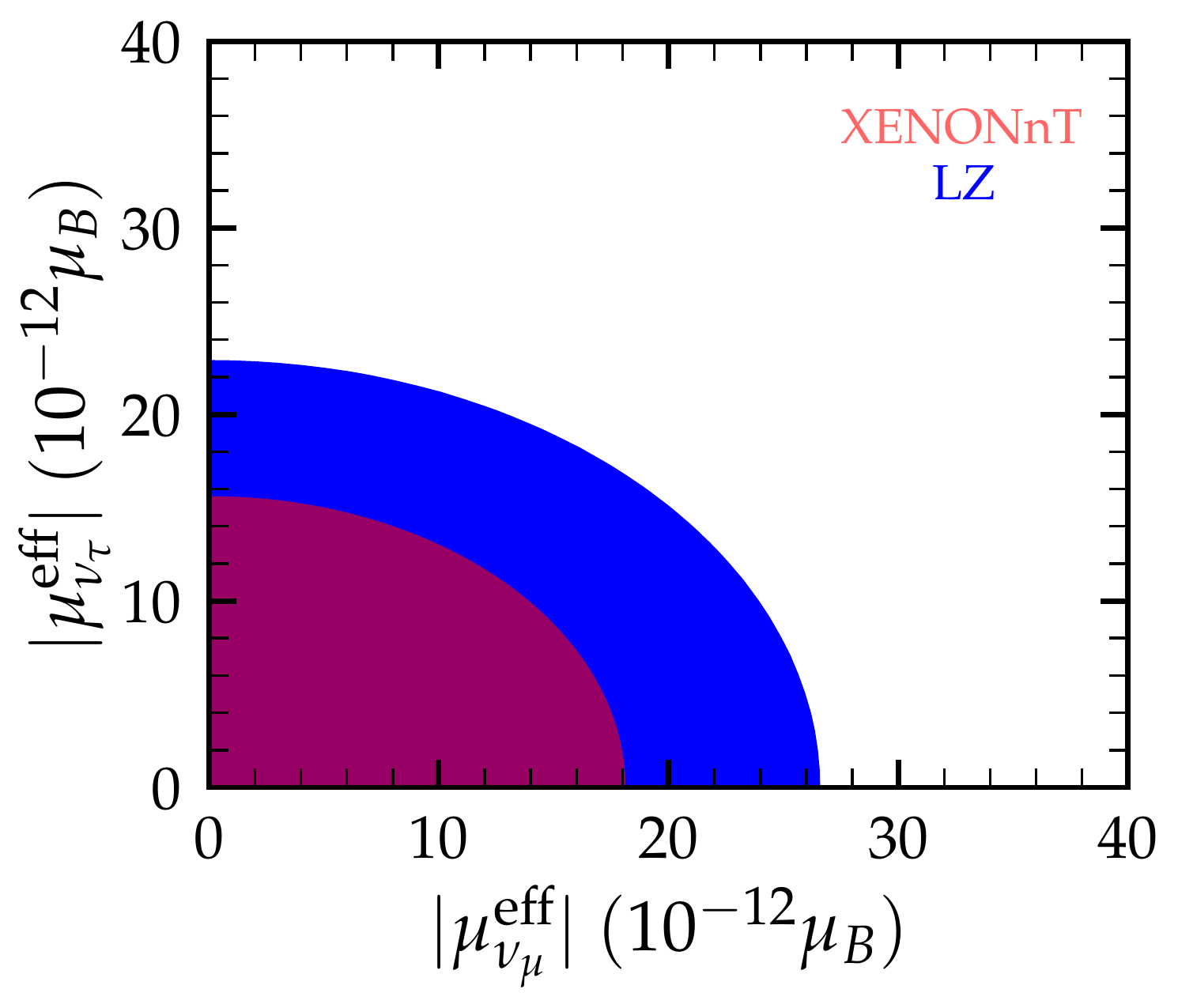}\\
 \includegraphics[width=0.32 \textwidth]{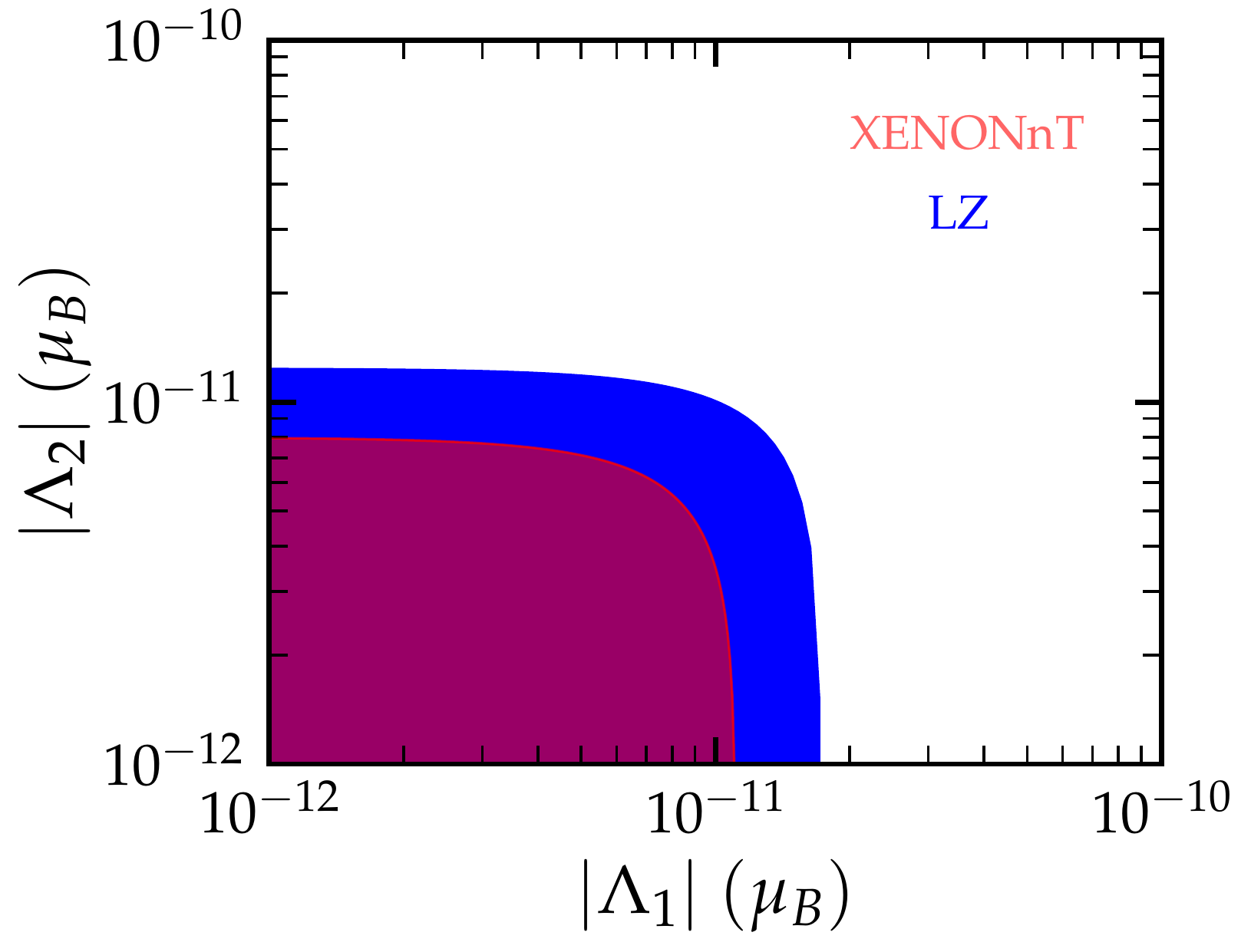}
 \includegraphics[width=0.32 \textwidth]{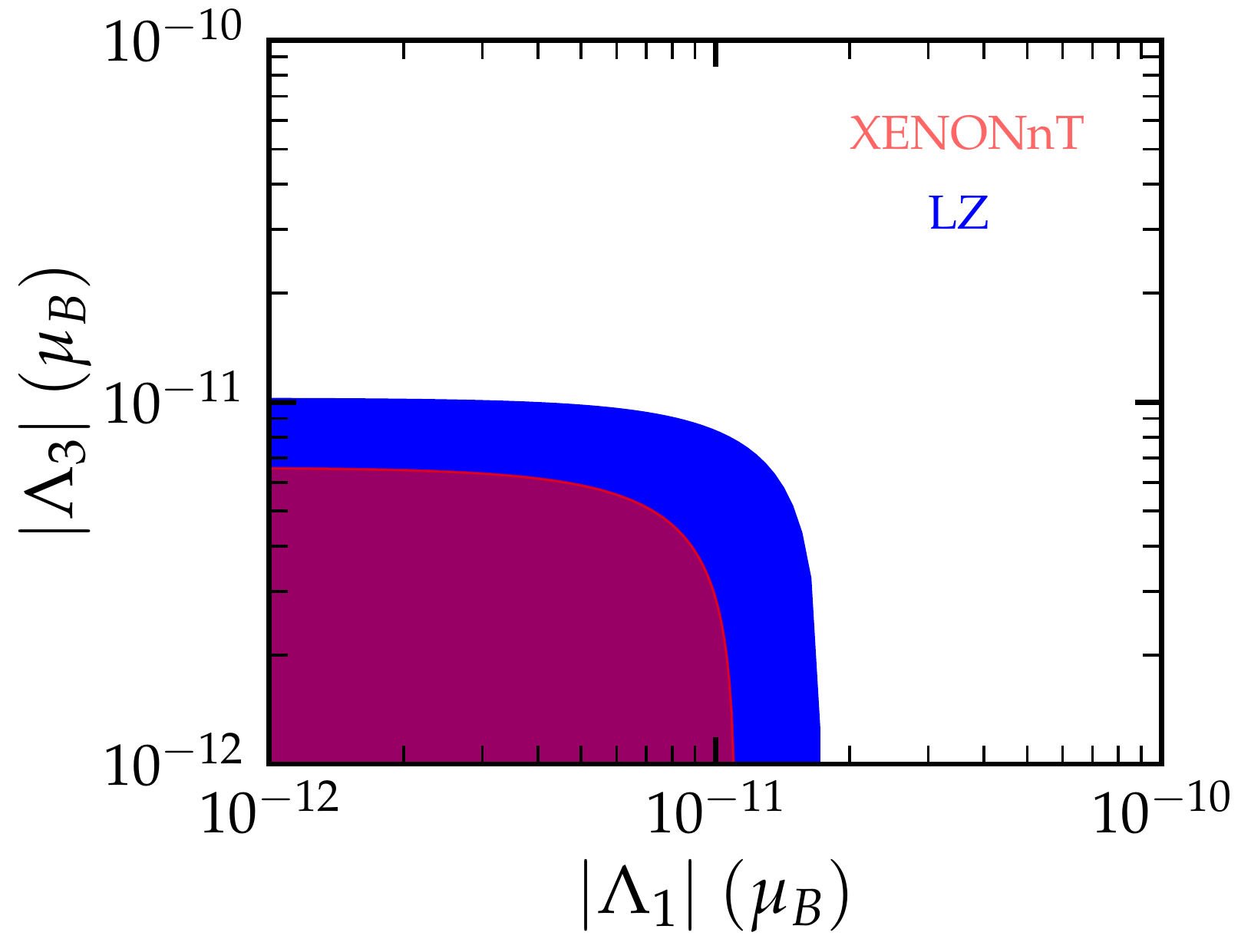}
 \includegraphics[width=0.32 \textwidth]{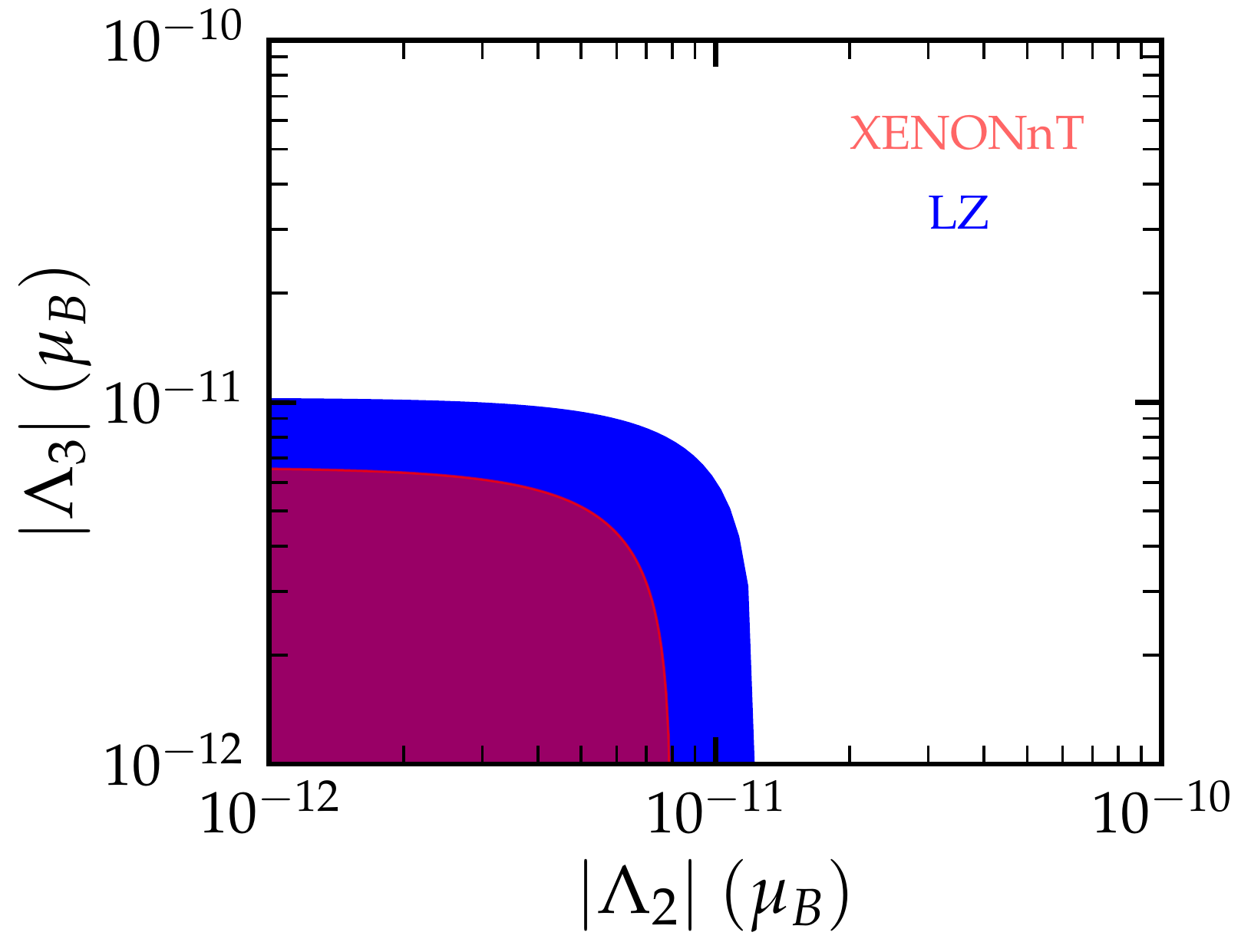}
 \caption{Regions allowed by LZ (blue) and XENONnT (red) data  at 90\% C.L. assuming two nonvanishing effective neutrino magnetic moments at a time (upper panel) and in the TMMs parameter space $(\Lambda_i, \Lambda_j)$ by marginalizing over the undisplayed parameter $\Lambda_k$ (lower panel).}
 \label{fig:mag_contours}
 \end{figure*}

In Fig.~\ref{fig:v_Electro_Magnetic_Properties} we show the one-dimensional $\Delta\chi^2$ profiles corresponding to the effective neutrino magnetic moment, obtained  from the analysis of LZ and XENONnT data. Differently from Ref.~\cite{AtzoriCorona:2022jeb} where a universal effective neutrino magnetic moment has been considered, here we present the individual limits on the flavored effective magnetic moments according to Eq.(\ref{eq:xsec_flavored}). At 90\% C.L. we find the upper limits: $$\{\mu_{\nu_e}^\text{eff},~\mu_{\nu_\mu}^\text{eff},~\mu_{\nu_\tau}^\text{eff}\}=\{13.9 (9.0),~22.8 (14.7),~19.6 (12.7)\}\times 10^{-12} \mu_B\, ,$$ for the case of LZ (XENONnT). The latter constitute  the most severe limits extracted from laboratory-based experiments to date, surpassing existing limits from the analysis of Borexino Phase-II data~\cite{Borexino:2017rsf} carried out in Ref.~\cite{Coloma:2022umy} for $\mu_{\nu_e}^\text{eff}$, $\mu_{\nu_\mu}^\text{eff}$ and $\mu_{\nu_\tau}^\text{eff}$ as well as the TEXONO~\cite{TEXONO:2006xds} and GEMMA~\cite{Beda:2013mta} limits on $\mu_{\nu_e}^\text{eff}$.  Assuming an effective neutrino magnetic moment that is universal over all flavors we find the upper limits: $\mu_\nu^\text{eff}=10.1~(6.3)\times 10^{-12} \mu_B$ for LZ (XENONnT). Notice, that the latter limit is in excellent agreement with the one reported by XENONnT \cite{Aprile:2022vux}. Going one step further, for the first time we derive the corresponding constraints on the fundamental magnetic moments  $\lambda_{ij}$ (see Section~\ref{sec:theory} and Refs.~\cite{Grimus:2002vb, AristizabalSierra:2021fuc} for details). Using the definition $\Lambda_i = \epsilon_{ikj} \lambda_{jk}$ we find the limits: $$ \{\Lambda_1, \Lambda_2, \Lambda_3\} =\{17.2 (11.1),~12.3 (8.0),~10.2 (6.6)\}\times 10^{-12}  \mu_B\, ,$$ at 90\% C.L. from the analysis of LZ (XENONnT) data. The latter limits are directly comparable and competitive with astrophysical limits derived from plasmon decay:  $\mu_{\nu}^\text{eff}=\sqrt{\sum_i |\Lambda_i|^2}=4.5\times 10^{-12}~\mu_B$ (95\% C.L.)~\cite{Viaux:2013hca}. In the upper panel of Fig.~\ref{fig:mag_contours} we show the 90\% C.L. allowed regions in the ($\mu^\text{eff}_{\nu_\alpha}, \mu^\text{eff}_{\nu_\beta}$) plain assuming the third effective magnetic moment to be vanishing, while in the lower panel we demonstrate the corresponding 90\% C.L. limits  in the TMM parameter space $(\Lambda_i, \Lambda_j)$ by marginalizing over $\Lambda_k$.

\begin{figure}[h]
\begin{center}
\includegraphics[width=0.49\textwidth]{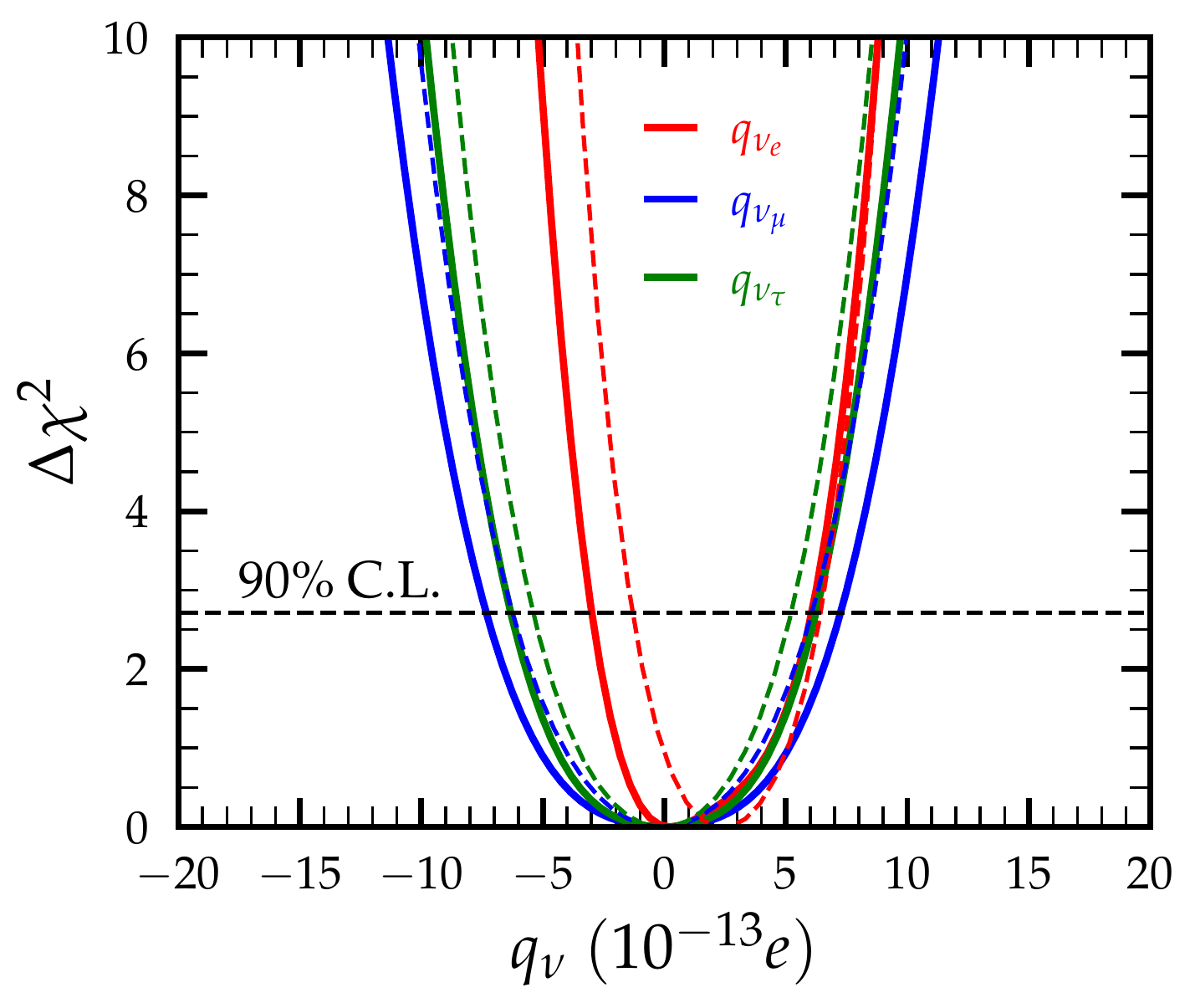}
\includegraphics[width=0.49\textwidth]{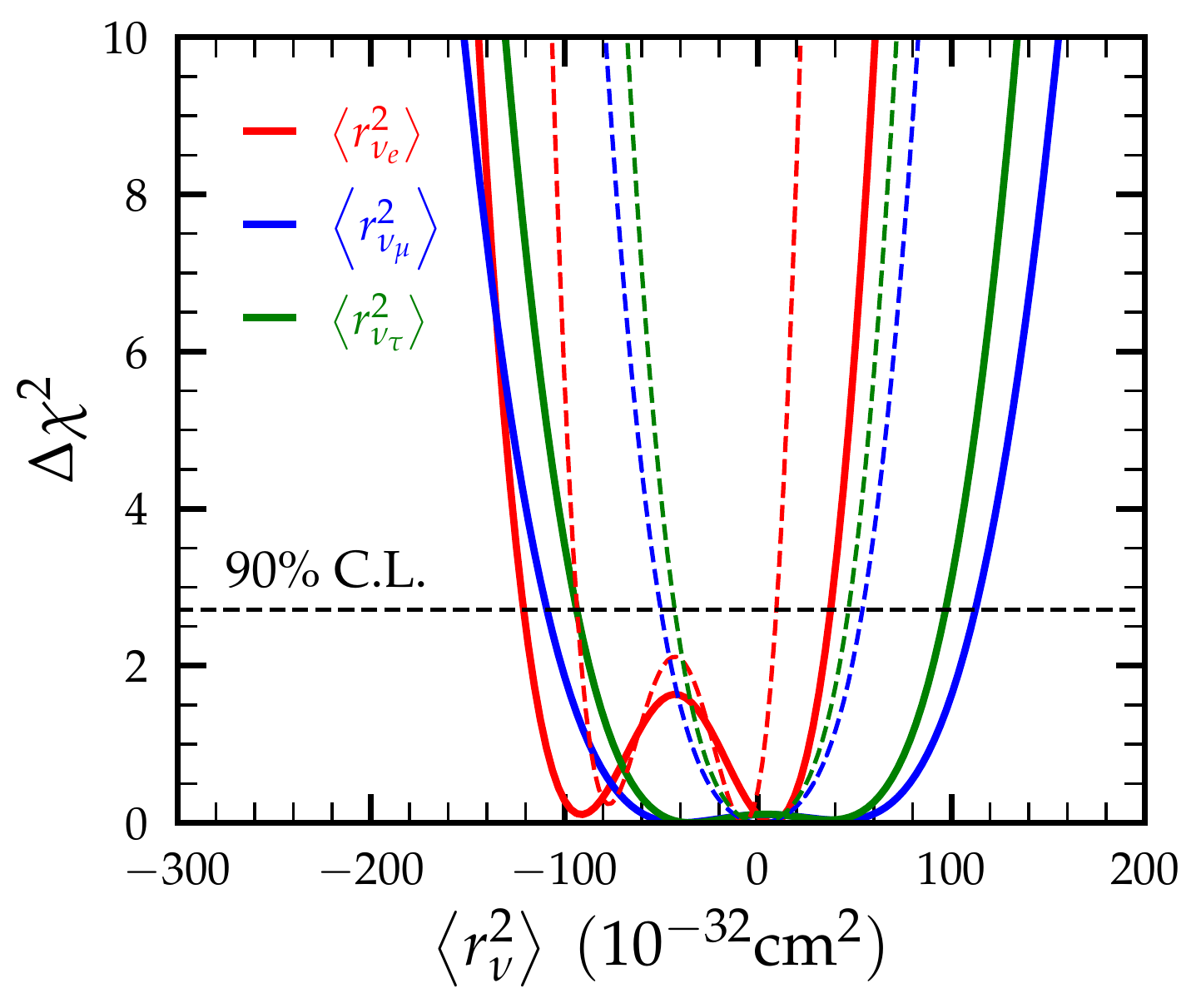}
\end{center}
\caption{$\Delta\chi^2$ profiles of the flavor dependent neutrino millicharge (left), charge radius (right). The results correspond to the analysis of LZ data (solid lines) and XENONnT data (dashed lines).}
\label{fig:v_Electro_Magnetic_Properties2}
\end{figure}
\begin{table*}[t]
\begin{adjustbox}{max width=\textwidth}
		\begin{tabular}{| c| c |c | c | c|}
			\hline
			\hspace{0.5cm}Flavor\hspace{0.5cm}   & \hspace{0.5cm} $\vert\mu_{\nu} \vert \; [10^{-11} \mu_B]$ \hspace{0.5cm} & \hspace{0.5cm} $q_{\nu} \; [10^{-12} e]$ \hspace{0.5cm}&$\hspace{0.5cm} \langle r^2_{\nu} \rangle \; [10^{-32} \text{cm}^2]$ \hspace{0.5cm} \\
			\hline \hline
			$\nu_e$ \ \     & $ \boldsymbol{\leq 1.4 \;}$ \textbf{(LZ)}\ \  &$\boldsymbol{[-0.3,0.6]}$\;(\textbf{LZ}) &$\boldsymbol{ [-121,37.5]}\;$(\textbf{LZ})  \\[0.1cm]

			 \ \     & $\boldsymbol{\leq 0.9 }$\; \textbf{(XENONnT)}\ \  &$\boldsymbol{[-0.1,0.6]}$\;(\textbf{XENONnT}) &$\boldsymbol{[-93.4,9.5]\;}$(\textbf{XENONnT})  \\[0.1cm]

			 \ \       & $\leq 3.7 \; (\text{Borexino})$~\cite{Coloma:2022umy} \ \      & $\leq 1\; \text{(Reactor)}$ \cite{Chen:2014dsa} &$[-4.2, 6.6]$ (TEXONO) \cite{TEXONO:2009knm} \\[0.1cm]
			
			 \ \             & $\leq 7.4\; \text{(TEXONO)}$~\cite{TEXONO:2006xds} \ \    &$[-9.3,9.5]\;\text{(Dresden-II)}$\cite{AtzoriCorona:2022qrf} &$[-5.94,8.28]$ (LSND) \cite{PhysRevD.63.112001}\\[0.1cm]
			
			 \ \             & $ \leq 2.9\; \text{(GEMMA)} $~\cite{Beda:2012zz} \ \      & & $\text{[-7.1,5]~(COHERENT + Dresden-II)}$~ \cite{AtzoriCorona:2022qrf} \\ [0.1cm]

			\hline \hline
			
			$\nu_{\mu}$ \ \             & $ \boldsymbol{\leq 2.3}$\;(\textbf{LZ})\ \      &$\boldsymbol{[-0.7,0.7]}$\; (\textbf{LZ}) &$\boldsymbol{[-109,112.3] }$\;(\textbf{LZ})\\[0.1cm]

			 \ \             & $ \boldsymbol{\leq 1.5 }$\;\;(\textbf{XENONnT})\ \      &$\boldsymbol{[-0.6,0.6]}$\; (\textbf{XENONnT}) &$\boldsymbol{[-50.2,54]}$\; (\textbf{XENONnT})\\[0.1cm]

			 \ \             & $\leq 5 \; (\text{Borexino})$ \cite{Coloma:2022umy}\ \      &$\leq 11~\text{(XMASS-I)}$~\cite{XMASS:2020zke} &$[-1.2,1.2]$ (CHARM-II) \cite{VILAIN1995115} \\[0.1cm]

			 \ \             & \ \      & &$\text{[-5.9,4.3]~~(COHERENT + Dresden-II)~}$\cite{AtzoriCorona:2022qrf} \\[0.1cm]

			\hline \hline
			
			$\nu_{\tau}$ \ \             & $\boldsymbol{\leq 2}$\; \;(\textbf{LZ})\ \      &$\boldsymbol{[-0.6,0.6]}$\;(\textbf{LZ}) &$\boldsymbol{[-93.7,97]}$\;\textbf{~(LZ)} \\[0.1cm]

			 \ \             & $ \boldsymbol{\leq 1.3 }$\;(\textbf{XENONnT})\ \      &$\boldsymbol{[-0.5,0.5]}$\; (\textbf{XENONnT}) &$\boldsymbol{[-43,46.8]}$\;\textbf{~(XENONnT)} \\[0.1cm]

			 \ \             & $\leq 5.9 \; (\text{Borexino})$ \cite{Coloma:2022umy}\ \      & $\leq 11~\text{(XMASS-I)}$~\cite{XMASS:2020zke} & \\[0.1cm]
			\hline\hline
		\end{tabular}
\end{adjustbox}
\caption{Summary of 90\% C.L. limits on EM neutrino parameters neutrino magnetic moment (in units $10^{-11}~\mu_B$), millicharge (in units $10^{-12}~e$), charge radius (in units $10^{-32}~\mathrm{cm^2}$) extracted in the present analysis (bold font) using the LZ and XENONnT data. For comparison, also shown are the existing limits from other experiment. }
\label{tab:Comparison_of_v_EM_params}
\end{table*}

In the left (right) panel of Fig.~\ref{fig:v_Electro_Magnetic_Properties2} we present the corresponding sensitivities on the neutrino millicharge (charge radius). As for the case of the neutrino magnetic moment, the extracted constraints refer to the different flavors and indicate that the LZ and XENONnT data are very sensitive to this EM parameter. For each flavor we find the limits from the XENONnT  at 90\% C.L. $$\{q_{\nu_e},~q_{\nu_\mu},~q_{\nu_\tau}\}=\{(-1.3,6.4),~(-6.2,6.1),~(-5.4,5.2)\}\times 10^{-13}~e\, ,$$ which are by one order of magnitude more severe than existing constraints in the literature i.e. from TEXONO~\cite{Chen:2014dsa} as well as from those extracted in Ref.~\cite{AtzoriCorona:2022qrf} through a combined analysis of the recent coherent elastic neutrino-nucleus scattering (CE$\nu$NS) data by COHERENT~\cite{COHERENT:2020iec,Akimov:2021dab} and Dresden-II~\cite{Colaresi:2022obx}. Notice, that the XENONnT limits are by a factor 1.2 more stringent in comparison to the LZ limits which are found to be $$\{q_{\nu_e},~q_{\nu_\mu},~q_{\nu_\tau}\}=\{(-3,6.1),~(-7.3,7.2),~(-6.3,6.2)\}\times 10^{-13}~e\, ,$$ at 90\% C.L..  

Table~\ref{tab:Comparison_of_v_EM_params} summarizes the 90\% C.L. limits on EM neutrino properties extracted in the present work from the analysis of the LZ and XENONnT data. Also listed are the corresponding limits on the neutrino charge radii, for which as expected the new data are not placing a competitive sensitivity. This is due to the absence of signal enhancement at low momentum transfer, unlike the cases of neutrino magnetic moment and millicharge. For the sake of comparison, also shown in the Table are the most stringent existing limits placed from the different experimental data mentioned above. 

\begin{figure*}[t]
 \includegraphics[width=0.32\textwidth]{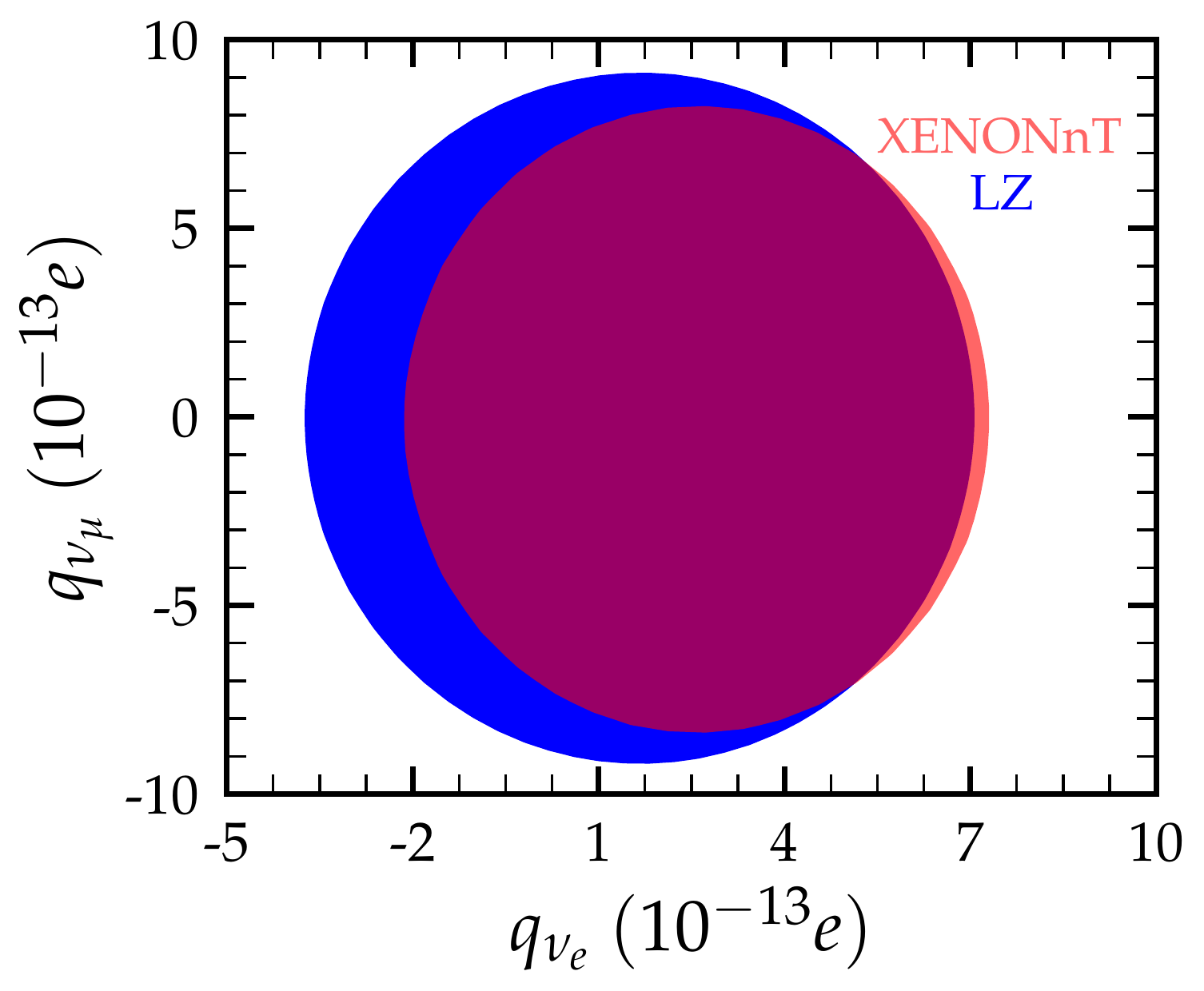}
 \includegraphics[width=0.32\textwidth]{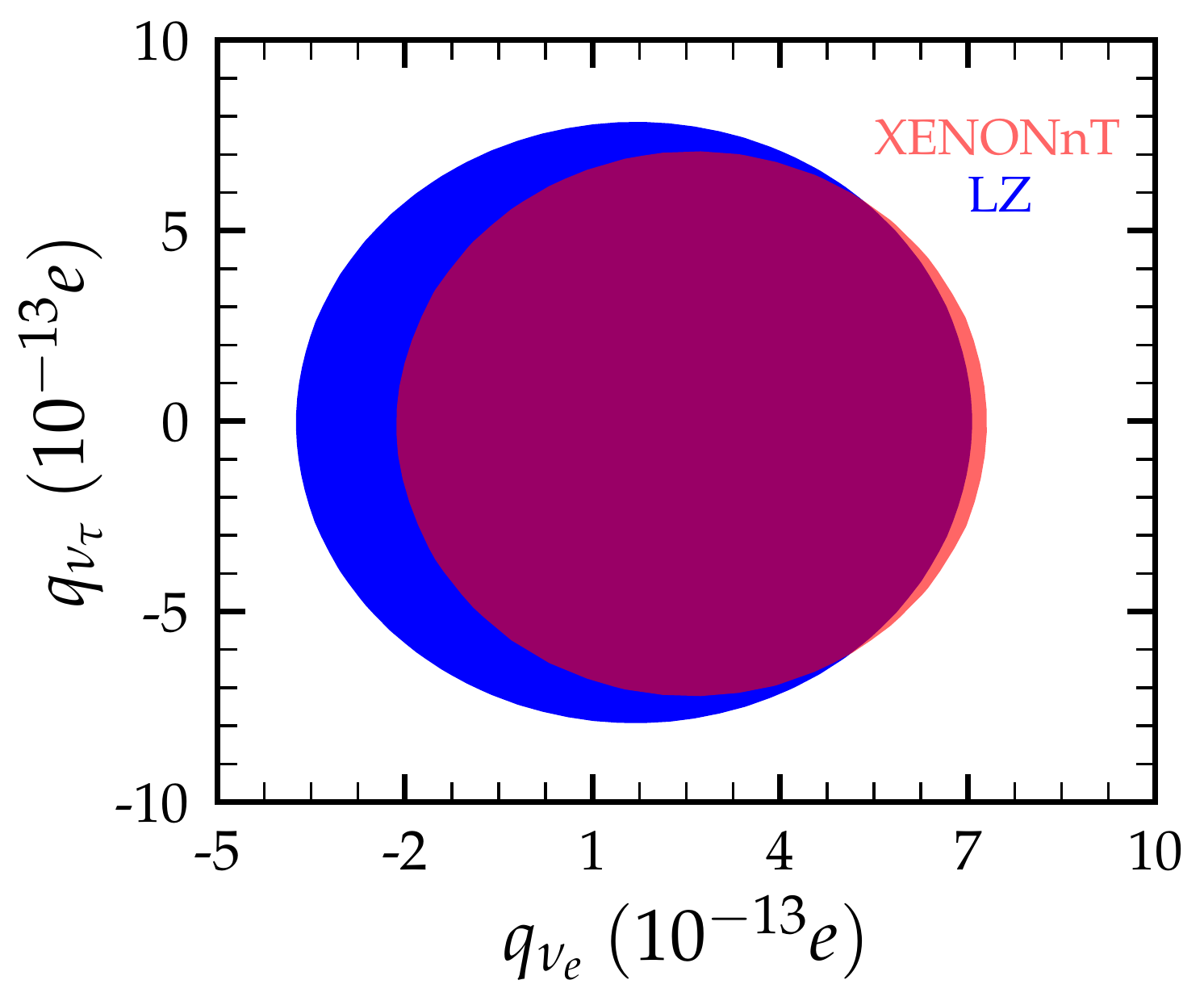}
 \includegraphics[width=0.32\textwidth]{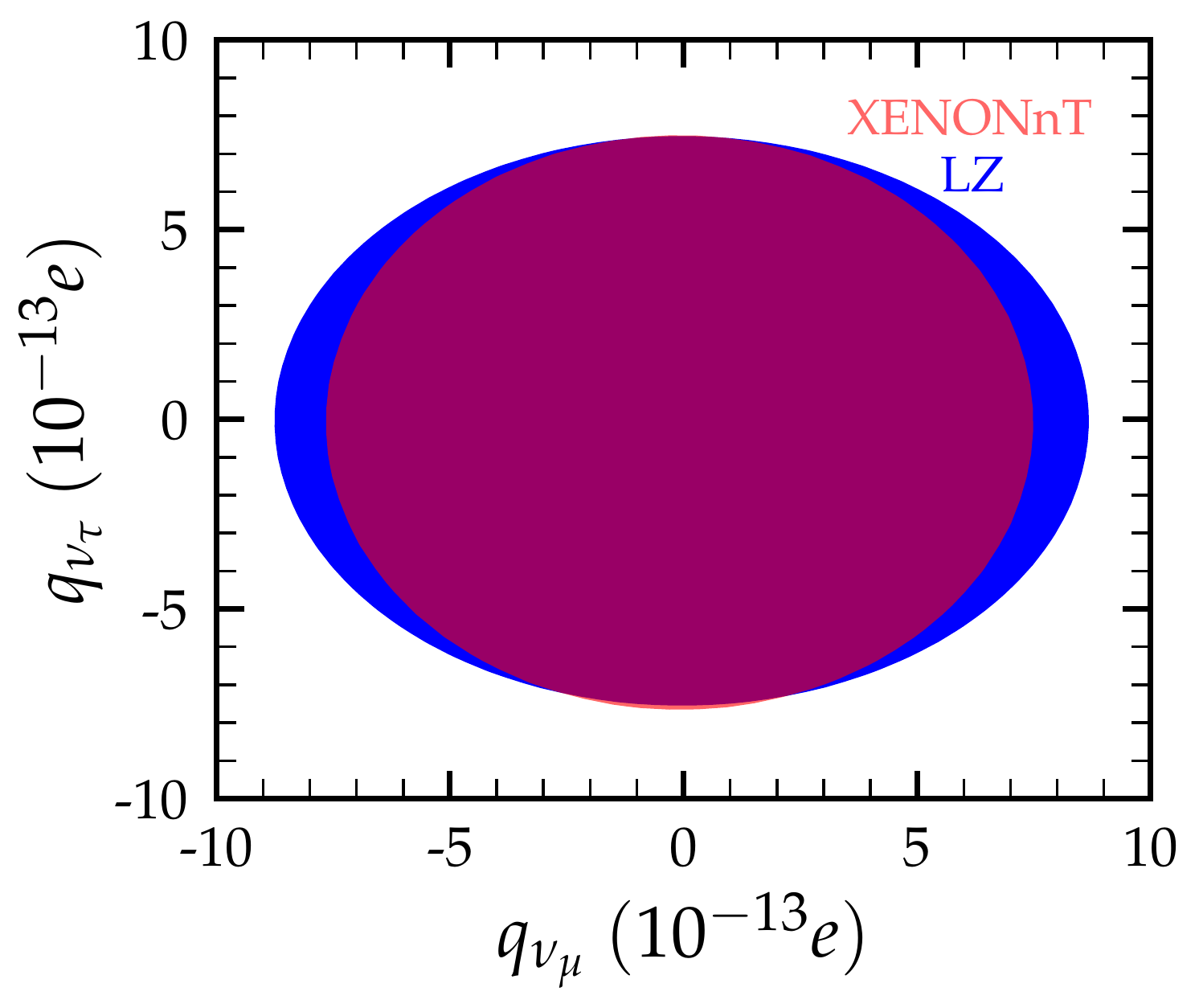}
 
  \includegraphics[width=0.32\textwidth]{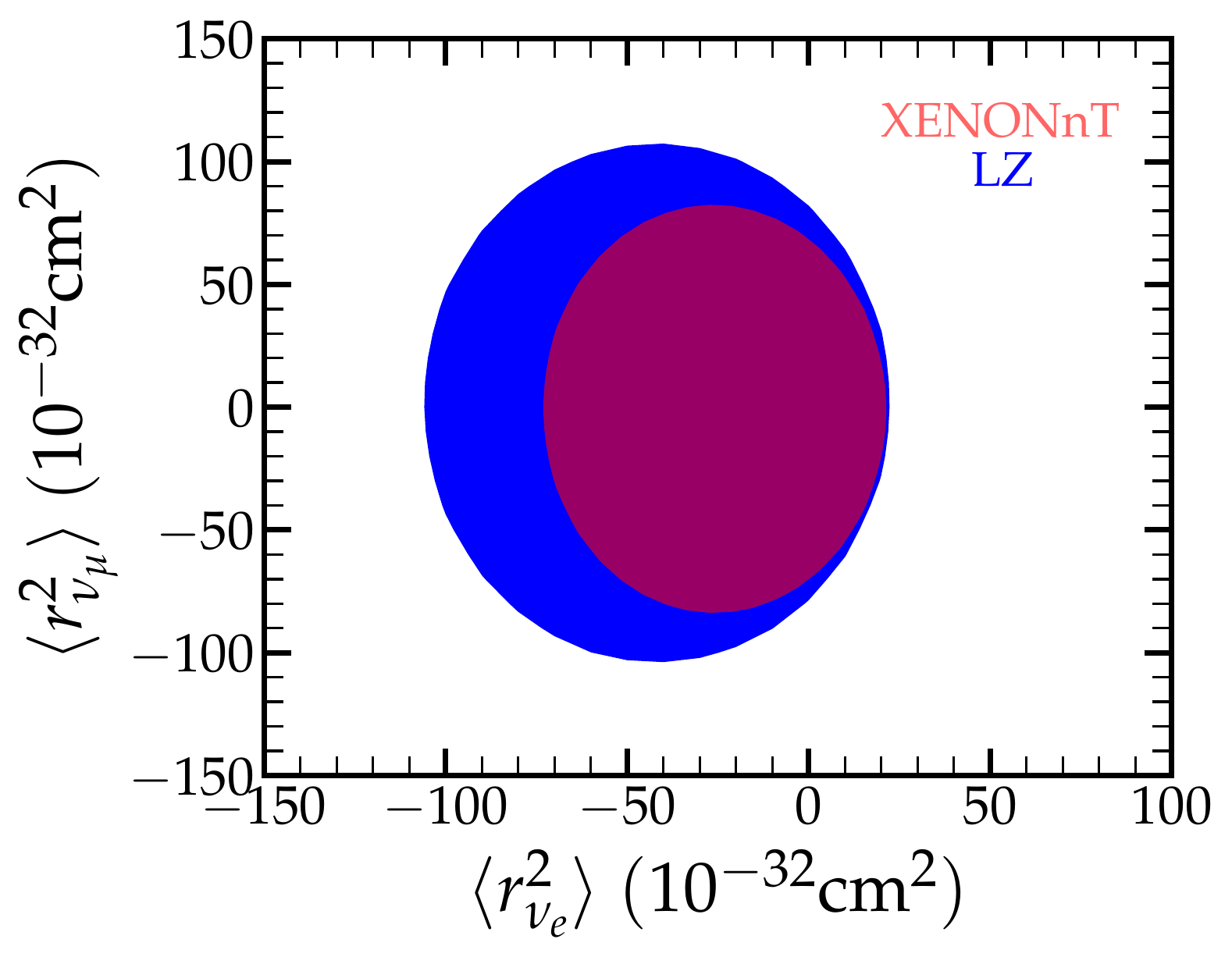}
 \includegraphics[width=0.32\textwidth]{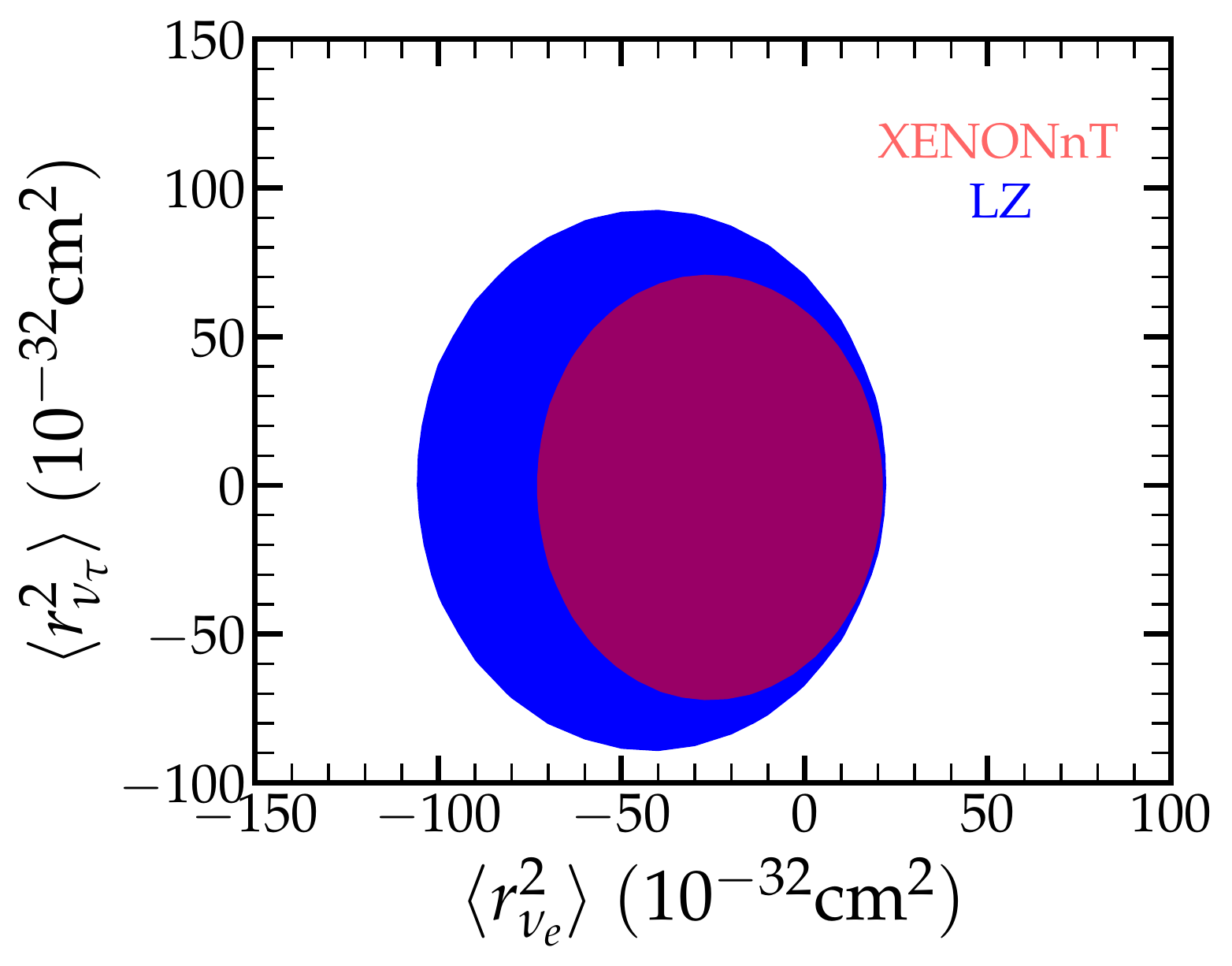}
 \includegraphics[width=0.32\textwidth]{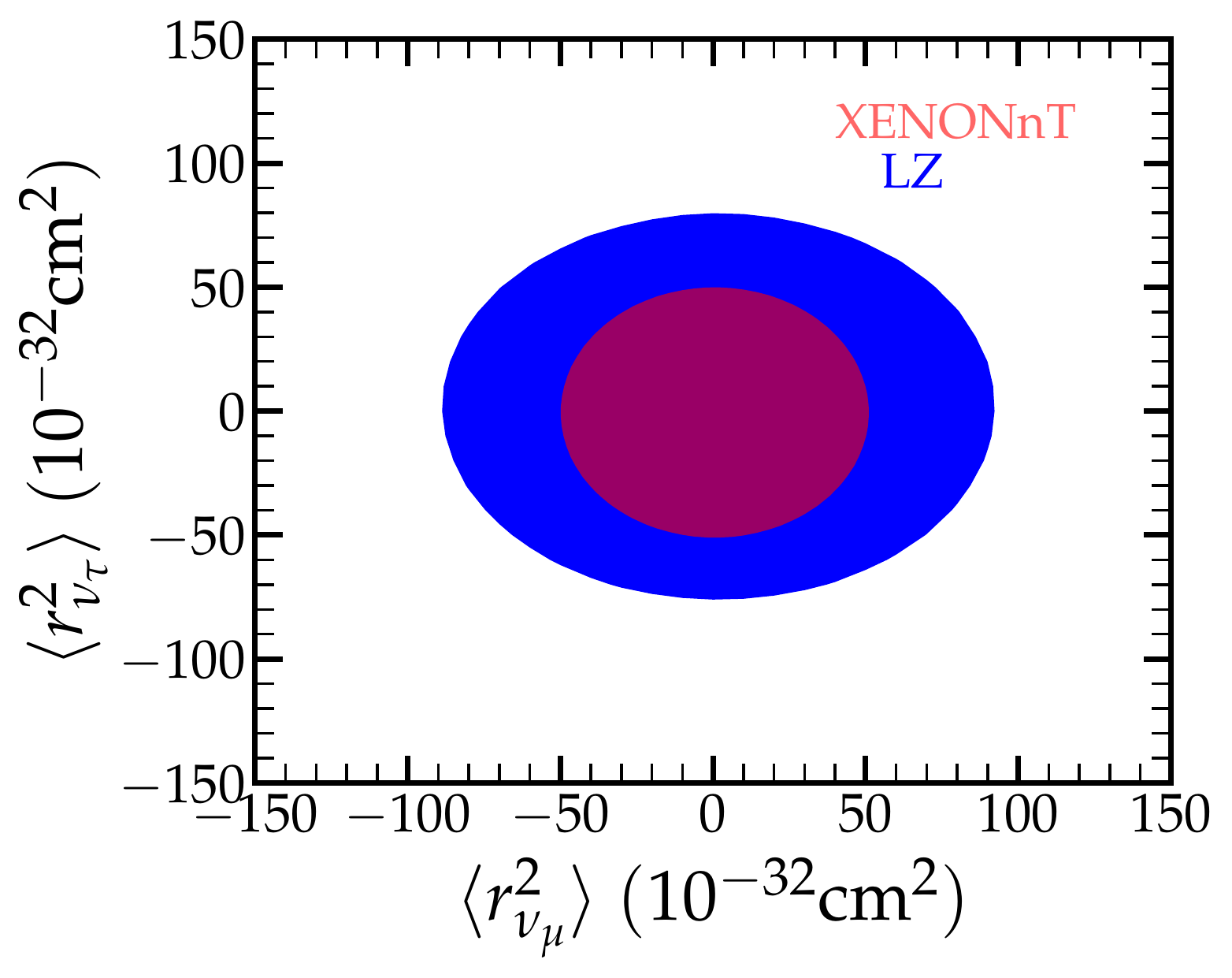}
 
 \caption{Upper panel: allowed regions in the millicharge parameter plains ($q_{\nu_\alpha}, q_{\nu_\beta}$). Lower panel: allowed regions in the charge radius parameter plains ($\langle r^2_{\nu_\alpha} \rangle, \langle r^2_{\nu_\beta} \rangle$). In both cases, the results are presented at 90\% C.L. and correspond to the analysis of LZ (blue) and XENONnT (red) data. Two nonvanishing parameters are allowed to vary at a time, while the third is set to zero.}
 \label{fig:milli_contours}
 \end{figure*}

At this point we are interested to perform a combined analysis allowing two nonzero neutrino parameters to vary at a time, assuming vanishing contribution from the third. First, for the case of neutrino millicharges the allowed regions at 90\% C.L. are illustrated in the upper panel of Fig.~\ref{fig:milli_contours}.
Then, assuming two nonzero charge radii at a time, the 90\% C.L. allowed regions in the parameter space of $(\langle r_{\nu_\alpha}^2 \rangle, \langle r_{\nu_\beta}^2 \rangle)$ are depicted in the lower panel of Fig.~\ref{fig:milli_contours}. Before closing this discussion, we wish to emphasize that LZ and XENONnT data can place only weak limits on the neutrino charge radii.

\begin{figure}[t]
\begin{center}
\includegraphics[width=0.49\linewidth]{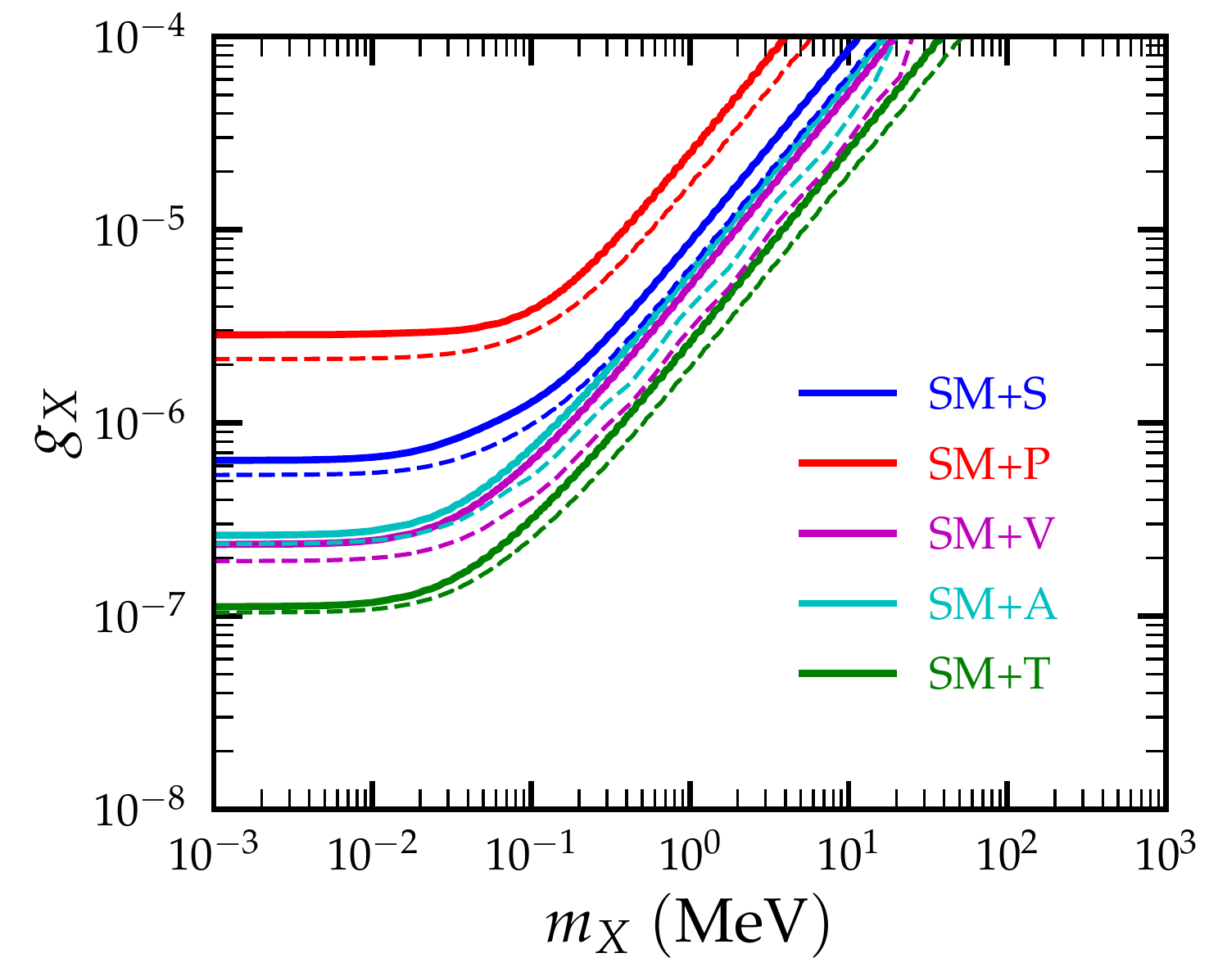}
\end{center}
\caption{90\% C.L. sensitivity in the NGI parameter space ($\textsl{g}_X$ and $m_X$) corresponding to the various interaction channels $X=\{S,P,V,A,T\}$, from the analysis of LZ (solid lines) and XENONnT (dashed) lines.}
\label{fig:v_NGI_paramters_space}
\end{figure}

\begin{figure*}[h!]
 \includegraphics[width=0.49\textwidth]{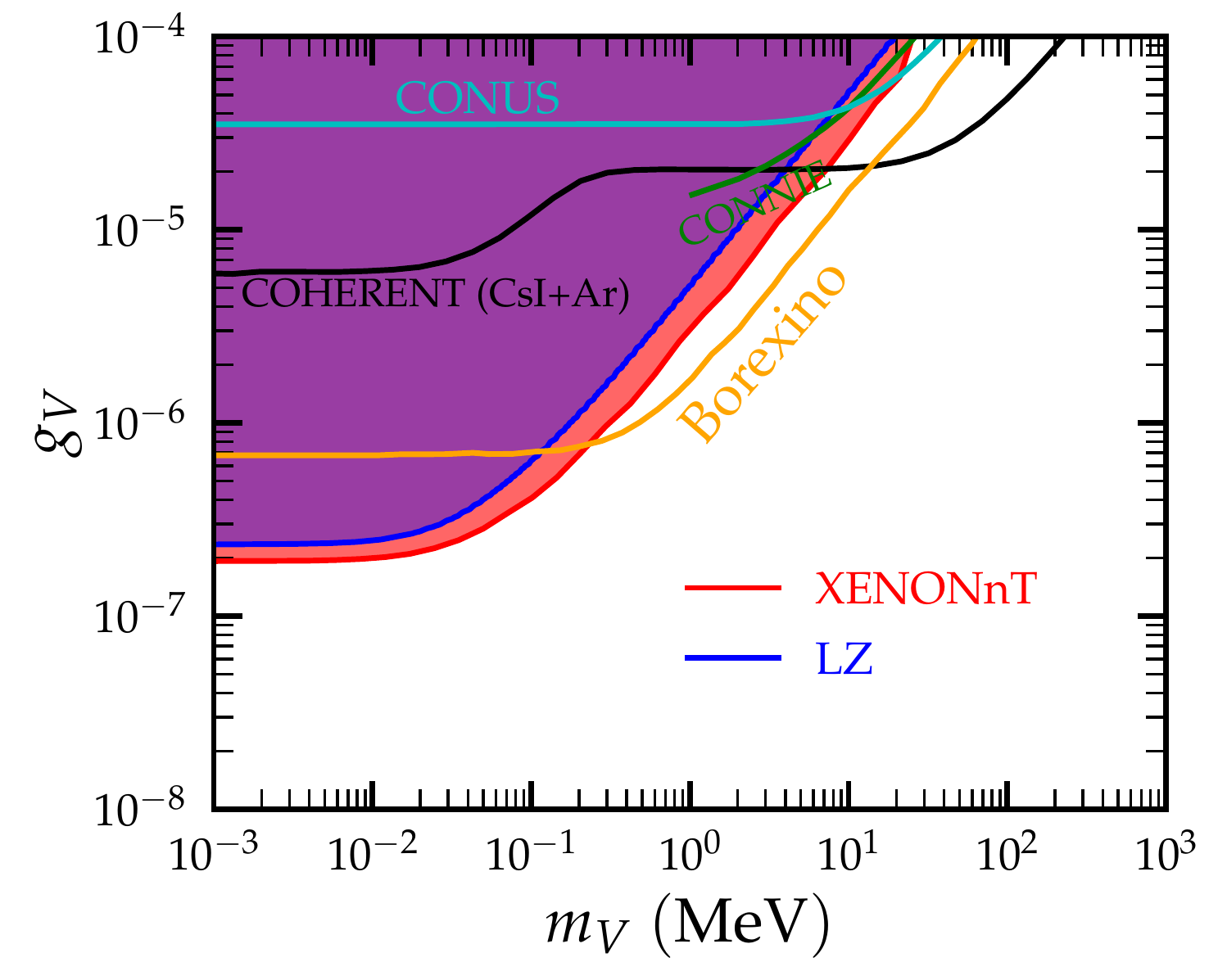}
 \includegraphics[width=0.49\textwidth]{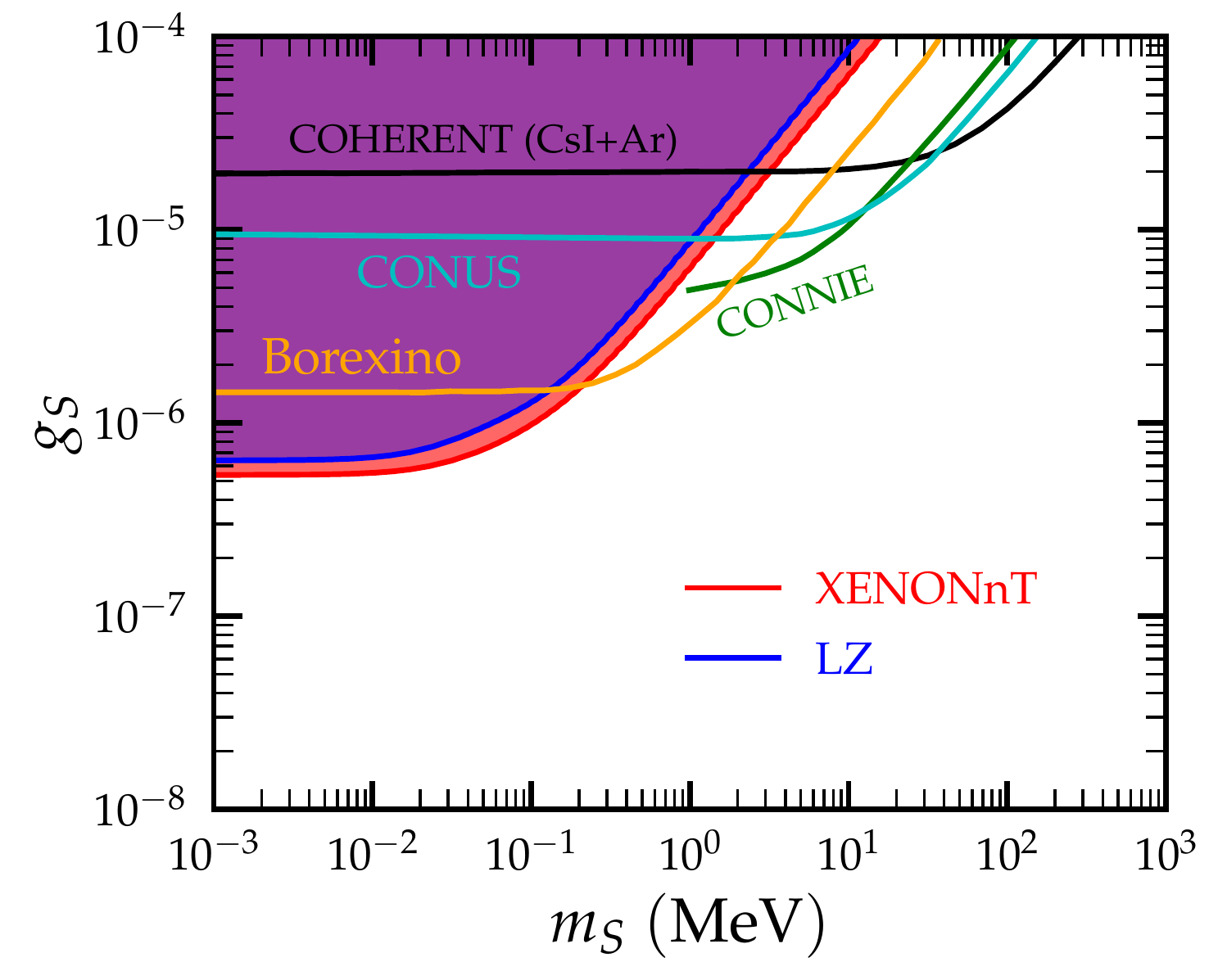}
 \caption{Sensitivity at 90\% C.L. for the the vector mediator model in ($m_V, \textsl{g}_V$) (left) and scalar mediator model in ($m_S, \textsl{g}_S$) (right). Existing constraints from other related studies are superimposed for comparison (see the text).}
 \label{fig:Scalar_Vector_Param_Space}
 \end{figure*}

We finally turn our attention on simplified NGIs with light $X=\{S,P,V,A,T\}$ mediators. The corresponding allowed regions by the LZ and XENONnT data  in the ($\textsl{g}_X,m_X$) plain are illustrated at 90\% C.L. in Fig.~\ref{fig:v_NGI_paramters_space}. We stress that NGI limits from the analysis of LZ data are presented for the first time in this work.   As can be seen, for the case of tensor (pseudoscalar) interaction the most (least) stringent bounds are found, in agreement with the projected sensitivities explored in Ref.~\cite{Majumdar:2021vdw}. In Fig.~\ref{fig:Scalar_Vector_Param_Space}, we first reproduce the limits from Fig.~\ref{fig:v_NGI_paramters_space} for vector (left panel) and scalar (right panel) NGIs.  Then, for a better comparison of our present work with existing limits in the literature, we superimpose limits coming from other laboratory experiments using \cevns data from: COHERENT~\cite{DeRomeri:2022twg} (See also Ref. \cite{AtzoriCorona:2022moj}),  CONNIE~\cite{CONNIE:2019xid} and CONUS~\cite{CONUS:2021dwh}, and through the \eves channel at Borexino~\cite{Coloma:2022umy}. Compared to Borexino Phase-II limits extracted in Ref.~\cite{Coloma:2022umy} the present analysis  leads to improved sensitivities, with the XENONnT data being slightly more constraining compared to LZ.

\section{\label{sec:conclusions}Conclusions}

Motivated by the low energy ROI as well as the high energy resolution and well-understood background at LZ and XENONnT, we have concentrated on new physics interactions characterized by signal enhancements at low momentum transfer that may lead to sizable signal distortions. In particular, we have analyzed the recent data reported by the two collaborations focusing on potential \eves contributions in the presence of EM neutrino properties and light NGI mediators. We find that in all cases the XENONnT data are competitive with LZ, though yielding slightly improved constraints. Since the two datasets lead to essentially similar sensitivities for the BSM scenarios considered, one would not expect a notable improvement from a combined analysis. Regarding the flavored effective neutrino magnetic moments as well as the fundamental transition magnetic moments, we show that the XENONnT data release set the currently best upper limits in the literature $\{9,14.7,12.7\}\times 10^{-12}~\mu_B$, being slightly more severe than the corresponding ones set by LZ and  improving  existing upper limits from Borexino and TEXONO. With respect to the neutrino millicharge, the present analysis leads to limits as low as $\{q_{\nu_e},~q_{\nu_\mu},~q_{\nu_\tau}\}=\{(-1.3,6.4),~(-6.2,6.1),~(-5.4,5.2)\}\times 10^{-13}~e$, hence improving previous upper limits by TEXONO and CE$\nu$NS experiments (COHERENT and Dresden-II) by one order of magnitude. On the contrary, we argue that LZ or XENONnT data are not sensitive to the neutrino charge radius (or the anapole moment). Finally, we conclude that for the case of simplified models accommodated in the framework of NGIs, the new data lead to improvements by about half an order of magnitude with respect to Borexino limits in the ($\textsl{g}_X,m_X$) parameter space.

\acknowledgments
\begin{justify}We thank the authors of Ref.~\cite{AtzoriCorona:2022jeb} for valuable discussions. DKP was supported by the Hellenic Foundation for Research and Innovation (H.F.R.I.) under the “3rd Call for H.F.R.I. Research Projects to support Post-Doctoral Researchers” (Project Number: 7036). The work of RS has been supported by the SERB, Government of India grant SRG/2020/002303.\end{justify}

\appendix
\section*{Comment on the anapole moment}
\begin{figure}[h]
\begin{center}
\includegraphics[width=0.49\textwidth]{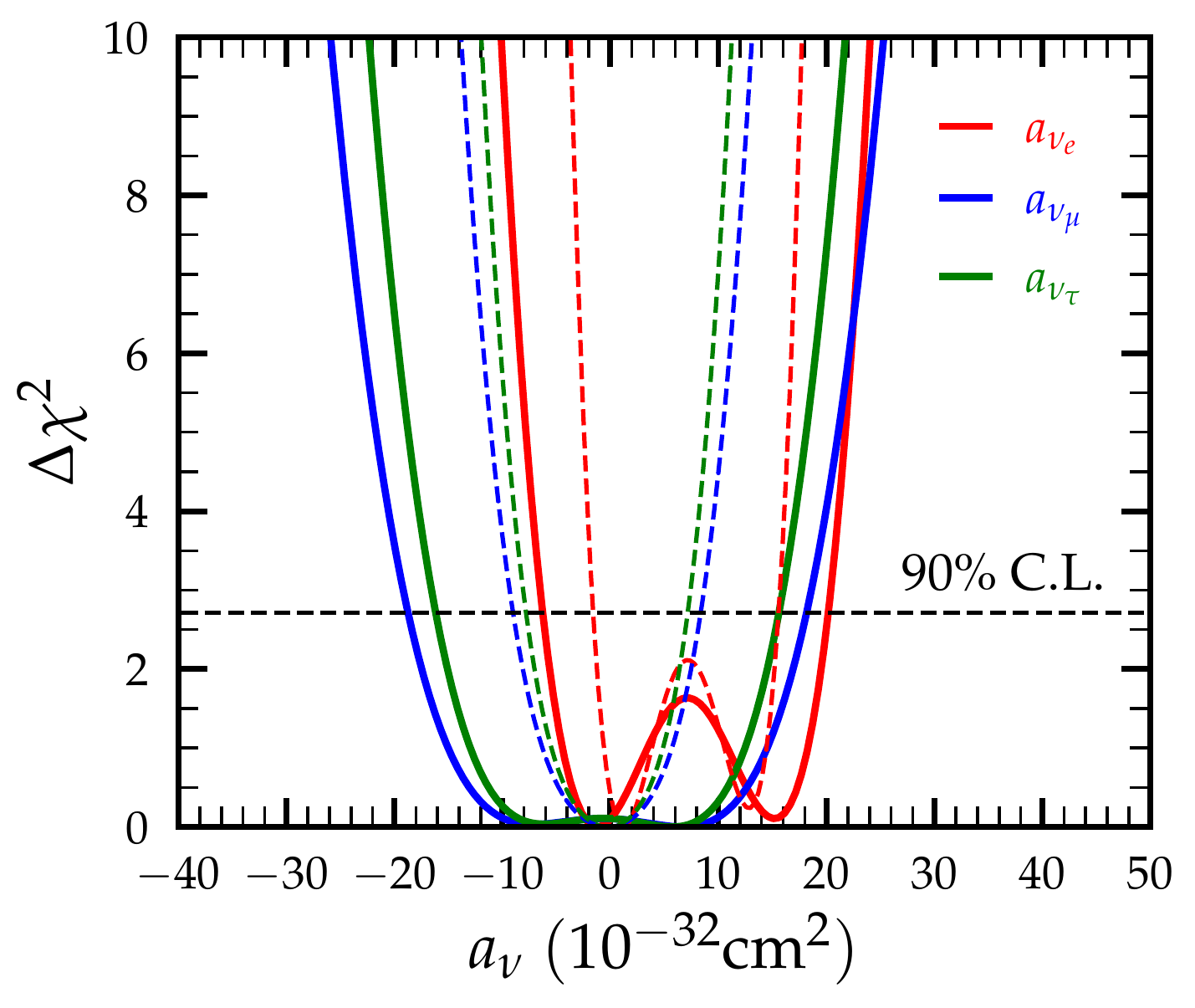}
\end{center}
\caption{$\Delta\chi^2$ profiles of the flavor dependent anapole moment. The results correspond to the analysis of LZ data (solid lines) and XENONnT data (dashed lines).}
\label{fig:v_Electro_Magnetic_Properties3}
\end{figure}

As pointed out in Ref.~\cite{Giunti:2014ixa} the charge radius and the anapole moment cannot be distinguished in the SM. Indeed, the anapole moment is related to the neutrino charge radius according to $a_{\nu_\alpha}= -\langle r^2_{\nu_\alpha} \rangle$/6. For completeness, in Fig.~\ref{fig:v_Electro_Magnetic_Properties3} we present the respective constraints on $a_{\nu_\alpha}$.

\bibliographystyle{utphys}
\bibliography{bibliography}

\end{document}